\documentclass[12pt]{article}
\usepackage{epsfig,amsfonts,amsthm}
\usepackage{latexsym}
\usepackage{amsmath}
\usepackage{amssymb}
\usepackage{bm}
\usepackage{color}

\newcommand{\be}{\begin{equation}}
\newcommand{\ee}{\end{equation}}
\newcommand{\bea}{\begin{eqnarray}}
\newcommand{\eea}{\end{eqnarray}}
\newcommand{\dst}{\displaystyle}
\newcommand{\fr}[2]{\frac{{\dst #1}}{{\dst #2}}}

\newcommand{\ba}{{\bf a}}

\newcommand{\bk}{{\bf k}}
\newcommand{\bK}{{\bf K}}

\newcommand{\lr}[1]{ \langle #1 \rangle}

\newcommand{\stolbik}[2]{ \left(\!\!\begin{array}{c}#1 \\ #2 \end{array}\!\!\right) }

\def\lsim{\mathrel{\rlap{\lower4pt\hbox{\hskip1pt$\sim$}}
    \raise1pt\hbox{$<$}}}         

\def\gsim{\mathrel{\rlap{\lower4pt\hbox{\hskip1pt$\sim$}}
    \raise1pt\hbox{$>$}}}         

\title{Elastic scattering of vortex electrons provides direct access to the Coulomb phase}

\author{I.~P.~Ivanov$^{1}$, D.~Seipt$^{2}$, A.~Surzhykov$^{3,4}$, S.~Fritzsche$^{2,5}$ 
\\
  {\small $^1$ CFTP, Instituto Superior T\'ecnico, Universidade de Lisboa, av.~Rovisco~Pais~1, 1049--001 Lisbon, Portugal}\\
  {\small $^2$ Helmholtz Institut Jena, D-07743 Jena, Germany}\\
  {\small $^3$ Physikalisch--Technische Bundesanstalt, D--38116 Braunschweig, Germany}\\ 
  {\small $^4$ Technische Universit\"at Braunschweig, D--38106 Braunschweig, Germany}\\ 
  {\small $^5$ Theoretisch-Physikalisches Institut, Friedrich-Schiller-Universit\"{a}t Jena, D-07743 Jena, Germany}\\
  }

\begin{document}

\maketitle
\begin{abstract}
Vortex electron beams are freely propagating electron waves carrying adjustable 
orbital angular momentum with respect to the propagation direction.
Such beams were experimentally realized just a few years ago and are now used to probe various electromagnetic processes.
So far, these experiments used the single vortex electron beams, either propagating in external fields
or impacting a target.
Here, we investigate the elastic scattering of two such aligned vortex electron beams and
demonstrate that this process allows one to experimentally measure features which are impossible
to detect in the usual plane-wave scattering. 
The scattering amplitude of this process is well approximated by two plane-wave scattering amplitudes
with different momentum transfers, which interfere and give direct experimental access
to the Coulomb phase.
This phase (shift) affects the scattering of all charged particles and
has thus received significant theoretical attention but was never probed experimentally.
We show that a properly defined azimuthal asymmetry, which has no counterpart
in plane-wave scattering, allows one to directly measure
the Coulomb phase as function of the scattering angle.
\end{abstract}

\section{Introduction}

Electron vortex beams (or twisted electrons) are electron states with helical wave fronts,
which carry non-zero orbital angular momentum (OAM) projection on the
average propagation direction \cite{review}.
Following the suggestion of \cite{bliokh-2007}, 
several groups recently reported production of vortex electrons with energies up to 300 keV
and an OAM as large as 100$\hbar$,
\cite{twisted-electron}. The state-of-the-art technology allows now to 
manipulate vortex electron beams \cite{thesis},
focus them to angstrom-size focal spots \cite{angstrom}, 
or to use them as a novel probe of various electromagnetic phenomena
such as an interplay of Larmor and Gouy rotation in longitudinal magnetic field \cite{probing}
or the acquisition of phase vortex in the field of an artificial magnetic monopole \cite{monopole}.
Several other proposals to use vortex electrons include the preparation
of structured beams \cite{structured-beams}, exploration of atomic transitions \cite{atomic-physics,SIFSS-2015},
and detection of unusual features of the electromagnetic radiation they emit 
\cite{transition-radiation,cherenkov-radiation}. These proposals still await experimental verification.

In all experiments conducted so far, however, the vortex electrons were impinging on a fixed target,
either a screen or a material specimen to be probed by vortex electrons.
One can readily envision that, by modifying the instrumentation, a ``collider'' of vortex electrons
can be formed in which the two vortex electron beams counterpropagate co-linearly and
are focused upon a common focal spot. 
Such collision of vortex electrons will lead to elastic scattering and other quantum-electrodynamical (QED) processes
and can be studied with conventional electron and photon detectors and spectrometers.
Because of the new degrees of freedom of vortex beams,
such collision experiments enable one to probe details of the scattering processes 
which the usual collisions are insensitive to \cite{ivanov-2011}.

There are publications which discuss scattering processes involving vortex electrons and other particles.
However, these authors considered either simplified processes with
``scalar'' vortex particles \cite{ivanov-2011,ivanov-2012} or Compton scattering in which only one
incident particle was twisted \cite{JS-2011,structured-beams}.
Here, we investigate the simplest QED scattering process,
M\o{ller} (elastic electron-electron) scattering, in which both initial electrons are twisted.
A brief report on this calculations appeared in \cite{short-paper}. 
We demonstrated there that this process serves as an analogue of the classical Young's 
double-slit experiment but in momentum space.
Just as in any interferometric technique, this experiment allows one to probe
the relative phase between the two ``momentum-space paths'', 
the two plane-wave scattering amplitudes with different momentum transfers.
For charged particle scattering, it allows one to measure
the momentum-transfer dependence of the Coulomb phase,
a quantity which has received significant theoretical attention 
but which has never been measured experimentally.

In this paper, we provide further details on these calculations, 
together with a qualitative as well as quantitative
numerical analysis of the results.
The structure of this paper is the following.
In the next Section, we calculate the M\o{}ller scattering of coaxial Bessel vortex electrons.
In particular, we highlight the all-important modifications to the transverse momentum distribution,
which reveal the intensity fringes arising from interference of two plane-wave scattering configurations.
Then, in Section~\ref{section-coulomb-phase} we demonstrate that this interference pattern
gets distorted by the momentum-transfer-dependent Coulomb phase,
and show how this dependence can be extracted from the measurements.
In Section~\ref{section-discussion}, we discuss the feasibility for detecting the interference and Coulomb phase
in such electron-electron collisions. We close the paper with a summary of our findings.
Several appendices contain supplementary material on vortex electrons 
and on $ee\to e'e'$ helicity amplitudes.

\section{M{\o}ller scattering of vortex electrons}\label{section-main}

\subsection{Notation and kinematics}\label{section-pureBessel}

Calculation of M{\o}ller scattering of Bessel electrons must be conducted within the fully relativistic framework.
To make the description self-contained, we start by introducing this formalism and applying it to Bessel electrons.
In this work, we use the definitions and conventions of \cite{SIFSS-2015};
other works, such as \cite{Bliokh-2011,Karlovets-2012}, use slightly different conventions.
Throughout the paper, we use relativistic Lorentz-Heaviside units: $\hbar = c = 1$, $e^2 = 4\pi\alpha_{em}$.
Bold letters correspond to transverse momenta with respect to the chosen $z$ axis (the beam axis), 
and the three-vectors are labeled with the vector symbol.

The plane-wave electron with the four-momentum
$k^\mu = (E,\, \bk,\, k_z)$,
where $\bk = |\bk|(\cos\phi_k,\,\sin\phi_k)$, $|\bk| = |\vec k| \sin\theta$, $k_z = |\vec k|\cos\theta$,
 and helicity $\lambda = \pm 1/2$ 
(the eigenvalue of the operator of the spin component along the electron momentum direction) is described by
\be
 \Psi_{k \lambda}(x)= {N \over \sqrt{2E}}\, u_{k \lambda}\,e^{-ikx}\,.
 \label{PW}
\ee
The bispinor $u_{k\lambda}$ used here is 
\be
u_{k\lambda} = \stolbik{\sqrt{E+m_e}\,w^{(\lambda)}}{2 \lambda \sqrt{E-m_e}\,w^{(\lambda)}}\,, \quad 
w^{(+1/2)} = \stolbik{c\, e^{-i\phi_k/2}}{s\, e^{i\phi_k/2}}\,,
\quad w^{(-1/2)} = \stolbik{-s\, e^{-i\phi_k/2}}{c\,e^{i\phi_k/2}}\,,\label{PWspinors}
\ee
where $c \equiv \cos(\theta/2)$, $s \equiv \sin(\theta/2)$.
The bispinors are normalized as 
$\bar u_{k\lambda_1} u_{k \lambda_2}= 2m_e\, \delta_{\lambda_1, \lambda_2}$
and $N = 1/\sqrt{V}$ is the normalization coefficient corresponding to one particle per large volume $V$.
We use this basis of plane-wave solutions of the Dirac equation to construct twisted electron, or the Bessel vortex state:
 \be
 \label{bessel}
 \Psi_{\varkappa m k_z \lambda}(x)= {N_{\rm tw} \over \sqrt{2E}}\, \int \fr{d^2 \bk}{(2\pi)^2}\,
  a_{\varkappa m}(\bk)\, u_{k \lambda}\,e^{-ikx},
  \quad N_{\rm tw}=\sqrt{\fr{\pi}{R L}},
  \ee
where the Fourier amplitude is
 \be
 a_{\varkappa m}(\bk)=(-i)^m \,e^{im\phi_k}\,
 \sqrt{\fr{2\pi}{\varkappa}}\,\delta(|\bk|-\varkappa)\,.
 \label{a}
 \ee
Here, the normalization coefficient $N_{\rm tw}$ differs from the plane-wave expression $N$
but still corresponds to one Bessel state electron per large cylindric volume $V=\pi R^2 L_z$. 
For a detailed discussion of how the twisted states look in the coordinate space
and how they must be normalized see \cite{JS-2011,ivanov-2011,Karlovets-2012,SIFSS-2015}. 

Notice that with the definition (\ref{bessel}) of the vortex electrons, we already fix a reference frame and 
the axis $z$. 
In particular, the wave function $\Psi_{\varkappa m k_z \lambda}(x)$ in Eq.~(\ref{bessel}) 
depends on this choice and is not Lorentz-invariant. It describes an electron 
that moves along axis $z$ with the longitudinal momentum
$k_z$, while its transverse motion is represented by a superposition of plane waves with
transverse momenta of equal modulus $\varkappa$ and of all azimuthal angles $\phi_k$. 
The so constructed Bessel electron state
possesses definite energy $E=\sqrt{\varkappa^2+k_z^2+m_e^2}$, definite helicity $\lambda$, 
as well as a well-defined value of the total angular momentum projection on the $z$ axis: $j_z=m$. 
Notice that $m$ is a half-integer number, \cite{SIFSS-2015}.

The orbital angular momentum and spin are not separately conserved due to the intrinsic spin-orbital interaction 
of the twisted electron, \cite{Bliokh-2011}.
However in the paraxial approximation, when $\theta \ll 1$, the spin-orbital interaction is suppressed.
If we neglect the spin-orbit interaction, we have two conserved quantum numbers: the $z$ projection of the spin $s_z$,
which in this approximation is equal to helicity $\lambda$,
and the $z$-projection of the orbital angular momentum $\ell = m - \lambda$.
One could also define Bessel electron states in which the spinor $u_{k\lambda}$ contains 
an extra factor $\exp(i\lambda\phi)$, while the Fourier amplitude (\ref{a}) is constructed with integer $\ell$
instead of half-integer $m$ \cite{Karlovets-2012}. This is also a valid Bessel electron solution; 
its total angular momentum depends on helicity, $j_z = \ell + \lambda$,
while the parameter $\ell$ characterizes the orbital angular momentum independent of helicity. 
The two conventions differ in how an unpolarized electron is defined;
in this work we will stick to the former definition.

To describe the collision of two aligned Bessel electrons, we take the first electron as in (\ref{bessel}) 
with all parameters carrying subscript 1,
while the second electron is constructed in a similar fashion with respect to {\em the same} $z$ axis 
but moving in the opposite direction.
The Fourier amplitude for the second electron $a_{\varkappa_2 m_2}(\bk_2)$ then contains the azimuthal factor $e^{-im_2\phi_2}$, 
because the azimuthal angle of any given plane-wave component $\phi_2$ is written in the chosen reference frame, 
which is $\pi$-rotated with respect to the ``native'' reference axis for that electron.

Performing a longitudinal boost, we find the frame in which the longitudinal momenta of initial electrons
are \textit{balanced}, i.e. $k_{2z} = - k_{1z}$, while the other parameters can still be different from each other: $m_2 \not = m_1$, 
$\varkappa_2 \not = \varkappa_1$, and therefore $E_2 \not = E_1$. 
The two final electrons in the elastic $ee\to e'e'$ scattering are described by plane waves
with four-momenta $k^{\prime \mu }_1$ and $k^{\prime \mu}_2$.
Their longitudinal momenta are also balanced, $k_{2z}' = - k_{1z}'$, and their energies satisfy $E_1' + E_2' = E_1+E_2$.
However, their final transverse momenta are not required to sum up
to zero or to any fixed vector,
because the initial electrons are not in a state of definite transverse momentum.
The only kinematical restriction 
is that the total final transverse momentum $\bK' = \bk_1' + \bk_2'$ lies within a ring 
defined by $\varkappa_1$ and $\varkappa_2$ \cite{ivanov-2011}:
\be
|\varkappa_1 - \varkappa_2| \le |\bK'| \le \varkappa_1 + \varkappa_2\,.\label{ring}
\ee
For such a scattering of two Bessel beams, the final phase space grows from
the single-particle angular distribution $d\Omega$ or the transverse momentum $d^2\bk_1'$
to the four-dimensional transverse momentum space $d^2\bk_1' d^2\bk_2' = d^2\bk_1' d^2 \bK'$.
As a consequence, further information can be extracted from the structures in the final kinematical distributions,
such as the $\bK'$-distribution at fixed $\bk_1'$.

If we select a kinematical configuration with final momenta $\bk_1'$ and $\bk_2'$, then the final energies 
are uniquely defined:
\be
E_1' = E_1 + \Delta E\,,\quad E_2' = E_2 - \Delta E\,, \quad 
\Delta E = {\bk_1^{\prime 2} - \bk_2^{\prime 2}\over 2(E_1 + E_2)} - {E_1 - E_2 \over 2}\,.
\ee
The final longitudinal momentum $k_{1z}' = - k_{2z}' = k_z'$
is also defined and can be calculated as 
\be
(k_z')^2 = (E_1 + \Delta E)^2 - \bk_1^{\prime 2} - m^2 = (E_2 - \Delta E)^2 - \bk_2^{\prime 2} - m^2\,.
\ee

\subsection{Scattering amplitude and cross section}

The high-energy particle scattering is usually calculated by assuming that the initial and 
final states are well approximated by plane waves.
The scattering matrix element for $ee\to e'e'$ plane-wave scattering with initial momenta $k_1$ and $k_2$
and final momenta $k_1'$ and $k_2'$ is then written as 
\be
S_{PW} = i(2\pi)^4\delta^{(4)}(k_1 + k_2 - k'_1 -k'_2)\cdot  
{{\cal M}(k_1,k_2;k_1',k_2') \over \sqrt{16 E_1 E_1' E_2 E_2'}}\cdot  N^4 \,, \label{SPW}
\ee
where the invariant amplitude ${\cal M}$ is calculated according to the standard Feynman rules
and where $N$ is the familiar plane-wave normalization coefficient.
Although in real experiment each initial particle is a wave packet centered at $\lr{k_i}$, the momentum
spread inside each of these packets is typically so small that the invariant amplitude ${\cal M}(k_1,k_2;k_1',k_2')$,
which is a smooth function of momenta,
can be taken constant and equal to ${\cal M}(\lr{k_1},\lr{k_2};k_1',k_2')$.
In this approximation we can split the transition probability unambiguously into the cross section and flux factors
\cite{KSS-1992}.

In this work we deal with initial states of essentially non-plane-wave nature,
and we aim at investigating how the invariant amplitude varies as a function of $k_1$ and $k_2$.
Therefore, we need to generalize the scattering amplitude to such situations.
In appendix~\ref{app-packet} we remind the reader of the general treatment of scattering 
of arbitrary monochromatic wave packets \cite{KSS-1992,ivanov-2012}.
In this Section we simplify this general theory by considering the two initial particles 
to be the Bessel vortex states. With the definition (\ref{bessel}), the $S$-matrix element can be written as
\bea
S &=& 
\int {d^2 \bk_1 \over (2\pi)^2} {d^2 \bk_2 \over (2\pi)^2} a_{\varkappa_1 m_1}(\bk_1)  a_{\varkappa_2 m_2}(\bk_2) 
S_{PW} {N_{\rm tw}^2 \over N^2}\nonumber\\[2mm]
&=& {i (2\pi)^4  \delta(\Sigma E) \delta(\Sigma k_z) \over \sqrt{16 E_1 E_1' E_2 E_2'}}N_{\rm tw}^2 N^2 
{(-i)^{m_1+m_2} \over (2\pi)^3\sqrt{\varkappa_1\varkappa_2}} \cdot {\cal J}\,,\label{Stw}
\eea
where $\delta(\Sigma E) \equiv \delta(E_1+E_2-E_1'-E_2')$, $\delta(\Sigma k_z) \equiv \delta(k_{1z}+k_{2z}-k_{1z}'-k_{2z}')$,
and the vortex amplitude defined as
\be
{\cal J} = \int d^2 \bk_1 d^2 \bk_2\, e^{im_1\phi_1 - im_2\phi_2}\,\delta(|\bk_1|-\varkappa_1) \delta(|\bk_2|-\varkappa_2)
\delta^{(2)}(\bk_1+\bk_2 - \bK')\cdot {\cal M}(k_1,k_2;k_1',k_2')\,.\label{J}
\ee
Squaring the $S$-matrix element, regularizing the squares of delta-functions as
\be
[\delta(\Sigma E)\delta(\Sigma k_z)]^2 = \delta(\Sigma E)\delta(\Sigma k_z) {T L \over (2\pi)^2}\,,
\ee
dividing the result by the total observation time, and integrating over the final longitudinal momenta,
we obtain the event rate:
\be
d\nu = {1 \over 16 E_1 E_2 (E_1+E_2) k_z'}\, L N_{\rm tw}^4
\, {|{\cal J}|^2 \over (2\pi)^6 \varkappa_1 \varkappa_2} \, d^2\bk_1' d^2 \bk_2'\,.\label{dnu0}
\ee
As explained in \cite{KSS-1992}, a splitting of the event rate into the differential cross section and 
the (conventional) luminosity can be unambiguously done only for plane waves,
while for non-plane-wave collisions it is a matter of convention. 
Following \cite{ivanov-2011}, we define here the flux according to
\be
j = (|\vec v_1| + |\vec v_2|)\int d^3 r |\psi_1(\vec r_1)|^2 |\psi_2(\vec r_2)|^2
= {k_z (E_1+E_2) \over E_1 E_2} L N_{\rm tw}^4 \cdot {\cal I}\,,
\ee
where $\vec v_1$ and $\vec v_2$ are the velocities of the colliding electrons and 
\be
{\cal I} = {\varkappa_1\varkappa_2 \over 2\pi}\int_0^R rdr [J_{m_1}(\varkappa_1 r)]^2 [J_{m_2}(\varkappa_2 r)]^2\,.\label{I-flux}
\ee
Here, $R$ is the same radius of the large but finite quantization volume
which was introduced in (\ref{bessel}).
This definition for the flux allows us to define the generalized cross section,
\be
d\sigma = {d\nu \over j} = {1 \over 16 (E_1+E_2)^2 k_z k_z'}\, {1 \over {\cal I}}
\, {|{\cal J}|^2 \over (2\pi)^6 \varkappa_1 \varkappa_2} \, d^2\bk_1' d^2 \bk_2'\,.\label{dsigma-tw}
\ee

The dynamics of the scattering process is determined by the vortex amplitude ${\cal J}$ defined in (\ref{J}).
It contains four integrations and four delta-functions, so that the integral can be done exactly. 
It is non-zero only if the inequality (\ref{ring}) is satisfied,
and there are only two points in the entire $(\bk_1,\,\bk_2)$ transverse momentum space which
contribute to its value.
\begin{figure}[h]
\centering
\includegraphics[width=0.75\textwidth]{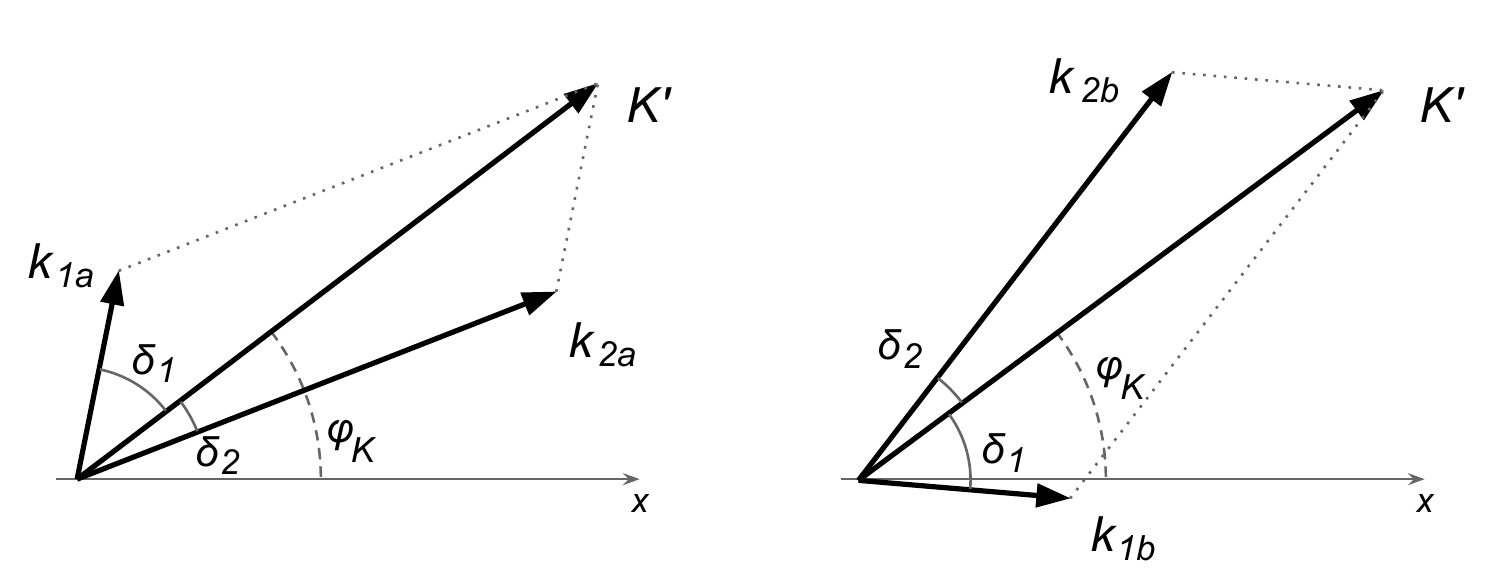}
{\caption{\label{fig-2configurations} The two kinematical configurations in the transverse plane that satisfy
momentum conservation laws in the scattering of two Bessel electron states.}}
\end{figure}
They correspond to the momenta $\bk_1$ and $\bk_2$ with the absolute values $\varkappa_1$ and $\varkappa_2$,
respectively, and the azimuthal angles
\bea
\mbox{configuration a:} &&\phi_1 = \phi_{K'} + \delta_1\,,\quad \phi_2 = \phi_{K'} - \delta_2\,,\nonumber\\
\mbox{configuration b:} &&\phi_1 = \phi_{K'} - \delta_1\,,\quad \phi_2 = \phi_{K'} + \delta_2\,.\label{phi12}
\eea
Here
\be
\delta_1 = \arccos\left({\varkappa_1^2 + K^{2} - \varkappa_2^2 \over 2\varkappa_1 K}\right)\,,
\quad
\delta_2 = \arccos\left({\varkappa_2^2 + K^{2} - \varkappa_1^2 \over 2\varkappa_2 K}\right)
\ee
are the internal angles of the triangle with the sides $\varkappa_1$, $\varkappa_2$, $|\bK'|$, 
and where for the sake of brevity we use $K = |\bK'|$.
These two kinematical configurations are shown in Fig.~\ref{fig-2configurations}.
The area of this triangle is
\bea
\Delta &=& {1 \over 2} K \varkappa_1 \sin\delta_1= {1 \over 2}K \varkappa_2 \sin\delta_2 
= {1 \over 2}\varkappa_1\varkappa_2\sin(\delta_1+\delta_2)\nonumber\\[1mm]
&=& {1 \over 4} \sqrt{2 K^2\varkappa_1^2 + 2 K^2\varkappa_2^2 + 2\varkappa_1^2\varkappa_2^2 
- K^4 - \varkappa_1^4 - \varkappa_2^4}\,.\label{area}
 \eea
The result for the vortex amplitude ${\cal J}$ can then be compactly written as \cite{ivanov-2011}
\be
{\cal J} = e^{i(m_1 - m_2)\phi_{K'}}{\varkappa_1 \varkappa_2 \over 2\Delta}
\left[{\cal M}_{a}\, e^{i (m_1 \delta_1 + m_2 \delta_2)} + {\cal M}_{b}\, e^{-i (m_1 \delta_1 + m_2 \delta_2)}\right]\,.\label{J2}
\ee
The plane-wave amplitudes ${\cal M}_{a}$ and ${\cal M}_{b}$ are calculated for the two distinct initial momentum 
configurations shown in Fig.~\ref{fig-2configurations} 
but for the same final momenta $k_1'$ and $k_2'$.

Finally, we need to insert the plane-wave amplitudes into Eq.~(\ref{J2}) 
to obtain the amplitude for the scattering of two Bessel electrons.
In the Born (one-photon exchange) approximation, the helicity amplitudes of M{\o}ller scattering are \cite{LL}
\be
{\cal M}_{\rm Born} = {\cal M}_t + {\cal M}_u = 4\pi\alpha_{em}\left({\bar u'_1 \gamma^\mu u_1\, \bar u'_2 \gamma_\mu u_2 \over t}
- {\bar u'_2 \gamma^\mu u_1\, \bar u'_1 \gamma_\mu u_2 \over u}\right)\,,
\ee
where each spinor is taken in the form (\ref{PWspinors}).
Here, $t = (k_{1} - k_1')^2 = (k_{2} - k_2')^2$
and $u = (k_1 - k_2')^2 = (k_2 - k_1')^2$ are the two relativistic invariant Mandelstam variables
which characterize the momentum transfer in two-particle scattering \cite{LL}.
They are different for the two interfering plane-wave amplitudes in (\ref{J2}):
\be
t_a - t_b = 2\bk_1' (\bk_{1a}-\bk_{1b}) = 4 |\bk_1'| \varkappa_1 \sin\delta_1 \sin(\phi_1' - \phi_{K'})\,.\label{tatb}
\ee
The third Mandelstam variable $s = (k_1+k_2)^2 = (k_1'+k_2')^2$ describes the total energy in the center of motion system 
squared and is the same for the two amplitudes.

In Appendix~\ref{app-helicity}, we give compact expressions for these helicity amplitudes.
Apart from the azimuthally-dependent kinematical factors, which can be reabsorbed in the definition
of initial and final wave functions, the Born amplitude is purely real.
Multi-photon exchanges modify this expression, and the result will be discussed in section~\ref{section-coulomb-phase}. 

\subsection{Bessel vortex electron scattering as a double-slit experiment in momentum space}

\begin{figure}[!htb]
\centering
\includegraphics[width=0.6\textwidth]{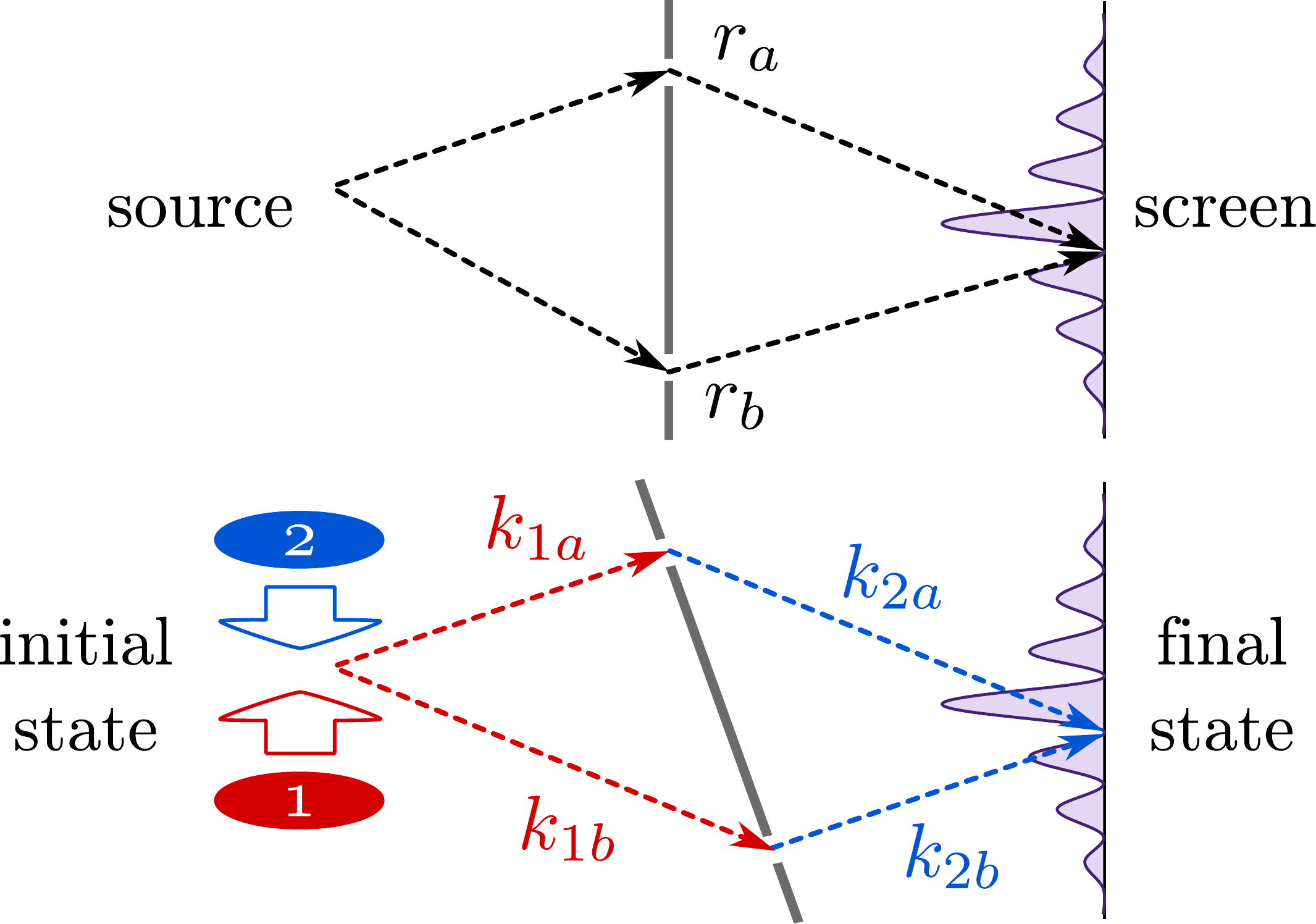}
{\caption{\label{fig-two-slit} Schematic illustration of the classic Young's experiment in coordinate space (upper image)
and of the double-slit experiment in momentum space (lower image). In the latter case, 
the arrows show that in the collision of two Bessel electron states (\ref{bessel}) 
only two momentum combinations lead to any final plane-wave state.}}
\end{figure}

Before we shall further analyze the cross section for the scattering of two Bessel beams, 
let us first look at the result (\ref{J2}) from a different perspective.
We argued in \cite{short-paper} that the elastic scattering of Bessel vortex electrons 
closely resembles the seminal Young's double-slit experiment but in momentum space.
In the usual double-slit experiment the wave emitted from a source propagates along 
two different spatial paths through two slits in a plate and interferes with itself on a distant screen, 
as shown in the upper image of Fig.~\ref{fig-two-slit}.
The superposition of the two amplitudes leads to a spatially periodic signal. 
Any change in the physical conditions along either path will be revealed by a shift 
of the interference pattern on the screen.

Eq.~(\ref{J2}) represents the momentum-space counterpart of this set-up.
The $ee\to e'e'$ scattering amplitude with pure Bessel vortex electrons in the initial state
is written as a sum of two plane-wave amplitudes ${\cal M}_a$ and ${\cal M}_b$, 
which interfere in the cross section and produce interference fringes.
This can be viewed as if the scattering process
evolves along two well-defined and well-separated ``paths'' in momentum space as schematically
shown in the lower image of Fig.~\ref{fig-two-slit}. The two momentum-space paths
end up in the same final state kinematics but the momentum transfers in each amplitude are different. 
The two plane-wave amplitudes are accompanied with phase factors 
which can be adjusted by selecting the final electron momenta.
By scanning the cross section across the allowed region of $k_1'$ and $k_2'$,
one observe the interference fringes in the final electrons' angular distribution.
This is analogous to the intensity stripes seen on a distant screen in the usual double-slit experiment.
The exact position and shape of the interference pattern 
is sensitive to the phase difference between ${\cal M}_{a}$ and ${\cal M}_{b}$,
and we will exploit this feature in Section~\ref{section-coulomb-phase}.

The interference pattern expected here is different from many other similarly-looking examples
of interference in collision experiments.
In most cases, an initial state evolves into a final state via {\em different intermediate states},
such as different excited states of an atom or different virtual particles in high-energy collisions.
For example, in neutrino oscillations \cite{neutrino} a neutrino is
produced in a state of definite flavour and propagates to the detection point as a superposition of 
three mass eigenstates. For a fixed neutrino energy, they correspond to different momenta,
and their interference causes spatially oscillating probability 
for changing flavour between production and detection point.
Although one can see it as the momentum-space
analogue of the two-slit experiment \cite{lipkin}, 
we stress that in this case the interfering amplitudes correspond to the same initial and final state kinematics 
but to different ``paths'' in the \textit{state space} in course of their propagation. 
In condensed matter physics, moreover, one encounters examples of interference 
between different momentum-space configurations of the same (quasi)particle along 
the same spatial path. Due to complicated dispersion law,
an electron with definite energy may have two different (quasi-) momenta which may interfere 
and may lead to a spatially varying electron density \cite{modulated-conductance}.
In this case, it is the medium that plays the instrumental role as it can absorb the extra momentum 
without destroying the coherence.

In contrast to these examples, elastic scattering of Bessel electron states
exhibits interference between two amplitudes with {\em the same state-space evolution} 
but with different combinations of momenta, and it 
arises in free space, without any medium to support the evolution.
Such examples were not known before.

\subsection{New dimension for the transverse momentum distributions}

In the usual $ee\to e'e'$ plane-wave scattering, 
the differential cross section contains the delta-function with the transverse momenta:
\be
d\sigma_{\rm PW} \propto \delta^{(2)}(\bk_1 + \bk_2 - \bk_1' - \bk_2') |{\cal M}|^2\, d^2\bk_1' d^2\bk_2' 
= |{\cal M}|^2 d^2\bk_1'\,.\label{dsigmaPW}
\ee
One can investigate the differential cross section in the $\bk_1'$-plane,
but their remains no freedom in choosing the transverse momentum of the second final particle
once the (transverse) momentum of the first final particle is fixed. 
In particular, in the center of motion frame, $\bk_2' = - \bk_1'$, and $\bK' = 0$.

\begin{figure}[h]
\centering
\includegraphics[width=0.5\textwidth]{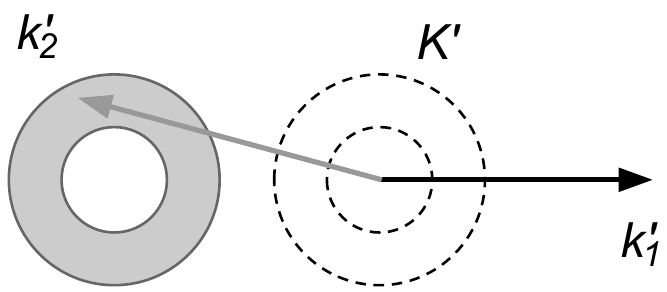}
{\caption{\label{figk1k2} At fixed transverse momentum $\bk_1'$, the allowed values of $\bk_2'$ fill the annular region centered
at $-\bk_1'$ (gray ring). The total transverse momentum $\bK'$ fills a similar ring (shown in dashed lines) around the origin.}}
\end{figure}

For Bessel vortex beam scattering, the cross section is given by Eq.~(\ref{dsigma-tw}).
The main difference, when compared to the plane-wave scattering, 
is that the angular distribution acquires a new dimension, 
as it depends explicitly on the transverse momenta of {\em both} final particles:
\be
d\sigma_{\rm tw} \propto |{\cal J}|^2 d^2\bk_1' d^2\bk_2' = |{\cal J}|^2 d^2\bk_1' d^2\bK'\,,\label{dsigmatw}
\ee
where ${\cal J}$ is given by (\ref{J2}).
This means, one can now study the $\bk_2'$-distribution or $\bK'$-distribution at {\em fixed} $\bk_1'$.
This distribution must lie inside the annular region shown in Fig.~\ref{figk1k2}.
This ring is centered at zero for the distribution on the $\bK'$-plane and 
at $-\bk_1'$ for the $\bk_2'$-plane, and the interference pattern resulting 
from the momentum-space double-slit experiment
will reveal itself as intensity stripes inside this ring.

Note also that this ring has a preferred orientation since the direction of $\bk_1'$
leads to a pattern of $(\bk_1',\,\bK')$-correlations.
It is convenient to quantify these correlations by means of the two azimuthal asymmetries:
\be
A_{\|} \equiv {\int d\sigma \cos(\phi_1' - \phi_{K'}) \over \int d\sigma}\,,
\quad
A_{\perp} \equiv {\int d\sigma \sin(\phi_1' - \phi_{K'}) \over \int d\sigma}\,.\label{asymmetries}
\ee
The integrals here can span either the entire space of momentum configurations
or a specific subregion inside the ring.
A non-zero and positive $A_{\|}$ indicates that $\bk_1'$ tends to dominate over $\bk_2'$;
a non-zero $A_\perp$ signals the loss of reflection symmetry
in the transverse plane with regard to the direction of $\bk_1'$.

\subsection{Angular distribution: qualitative discussion}\label{section-qualitative}

Before moving to numerical results, 
let us develop some intuition for the transverse momentum distribution in the simple case of
ultrarelativistic small angle scattering. In this case, $m_e\ll E$, $|t| \ll s$, 
and the $u$-channel contribution ${\cal M}_u$ can be neglected \cite{LL}.
The polar angles of the initial and final particles are
\be
\theta_1 \approx {\varkappa_1 \over k_z}\,,\quad \pi - \theta_2 \approx {\varkappa_2 \over k_z}\,,
\quad
\theta'_1 \approx {|\bk'_1| \over k'_z}\,,\quad \pi - \theta'_2 \approx {|\bk'_2| \over k'_z}\,,
\ee
which means that $c_1,\, c'_1,\, s_2,\, s'_2 \approx 1$, while $s_1,\, s'_1,\, c_2,\,c'_2$ are small.
The fermion helicity is conserved in the ultrarelativistic limit, so that the helicity amplitude can be written as:
\be
{\cal M} = 8 \pi \alpha_{em} {s \over t} e^{-i\lambda_1 (\phi_1-\phi_1')} e^{i\lambda_2 (\phi_2-\phi_2')}
\delta_{\lambda_1 \lambda_1'} \delta_{\lambda_2 \lambda_2'}\,.\label{Born-UR}
\ee
Substituting these amplitudes into (\ref{J2}), taking into account that configurations $a$ and $b$ correspond to
specific initial azimuthal angles (\ref{phi12}), one obtains the vortex amplitude squared for the unpolarized case 
\be
{1 \over 4}\sum_{\lambda_i} |{\cal J}|^2 = 64 \pi^2 \alpha^2_{em} s^2
{\varkappa_1^2\varkappa_2^2 \over 4\Delta^2}\left[{1 \over t_a^2} + {1 \over t_b^2} 
+ {2 \over t_a t_b}\cos(2m_1\delta_1+2m_2\delta_2)\cos\delta_1\cos\delta_2\right]\,.\label{Jsquared}
\ee
When deriving the above expression, we defined the unpolarized vortex electron as an incoherent
superposition of vortex states with a fixed value of the $z$-projection 
of total angular momentum $m$ and opposite helicities $\lambda$.
This convention is not unique, as we described above. 
One can also define the unpolarized electron by fixing
the orbital angular momentum $\ell = m-\lambda$, which is conserved in the paraxial limit. 
With this definition, the last term in the brackets of (\ref{Jsquared}) becomes
$\cos(2\ell_1\delta_1 + 2\ell_2 \delta_2)$. Which convention is more appropriate eventually depends
on the details of experiment, but their difference is not essential for the problem we consider.

Let us now analyze the behaviour of the vortex amplitude squared $|{\cal J}|^2$ inside the annular $\bK'$ region.
Eq.~(\ref{Jsquared}) shows that it depends on $K=|\bK'|$ via quantities $\delta_1$, $\delta_2$
and leads to a concentric ring structure.
These are the interference fringes characteristic of the Young's two-slit experiment 
but which now appear in the momentum space.
Of course, the particular number of stripes in this interference pattern depend on the values of $m_1$ and $m_2$.
Due to the $1/\Delta^2$ factor in (\ref{Jsquared}), the cross section diverges near the borders of the annular region,
where $\delta_1, \delta_2 \approx 0$ or $\pi$. 

Apart from dependence on $\delta_1$, $\delta_2$, the angle-differential cross section 
is also sensitive to the momentum transfer squared $t_{a,b}$. This dependence
is not azimuthally symmetric, as it involves all four transverse vectors,
which can be visualized by combining Figs.~\ref{fig-2configurations} and \ref{figk1k2}.
As the result, we expect a non-zero asymmetry parameter $A_\|$. To estimate its value, 
we assume further that $|\bk_i'| \gg \varkappa_i$.
Then, 
\be
{1 \over t_{a,b}} \approx -{1 \over \bk_1^{\prime 2}}\left[1 + {2\varkappa_1 \over |\bk'_1|}\cos(\phi'_1 - \phi_{K'} \mp \delta_1)\right]\,.
\ee
Grouping the factors in Eqs.~(\ref{dsigma-tw}) and (\ref{Jsquared}) that do not depend on the final-state kinematics
into the coefficient $C$, we express the Born-level cross section in this approximation as
\be
{d\sigma_{\rm Born} \over d^2\bk_1' d^2\bK'} = {C \over \Delta^2} {2 \over \bk_1^{\prime 4}}
\left[1 + {4\varkappa_1 \over |\bk'_1|}\cos\delta_1 \cos(\phi'_1-\phi_{K'})\right]
\bigl[ 1+ \cos(2m_1\delta_1+2m_2\delta_2)\cos\delta_1\cos\delta_2\bigr]\,.\label{J2tw2}
\ee
The overall effect is that the cross section increases when $\bK'$ is aligned with $\bk_1'$,
so that one can estimate $A_\| \sim \varkappa_1/|\bk'_1|$.
However, there is no up-down asymmetry on this plane: $A_\perp = 0$. 
This is due to the absence of terms proportional to $\sin(\phi'_1-\phi_{K'})$
or its odd powers. 

\subsection{Numerical results for pure Bessel electrons}

To corroborate the above qualitative analysis, the left panel of Fig.~\ref{fig-K-plots} shows
the numerical results for the differential cross section in the $\bK'$-plane for the following set of parameters:
\be
E_1 = 2.1\,\mathrm{MeV}\,, \ \varkappa_1 = 200\,\mathrm{keV}, \ \varkappa_2 = 100\,\mathrm{keV}, \ 
|\bk_1'| = 500\, \mathrm{keV}, \ m_1 = 1/2,\ m_2 = 13/2\,.\label{parameters}
\ee
All other kinematic parameters can be calculated from these input numbers. 
In this plot, $\bk_1'$ is directed to the right. 
As seen from this figure, both effects, the interference fringes and the correlation between $\bK'$ and $\bk_1'$
are readily observed in the cross sections. The number of interference fringes depends
not only on the values of $m_i$ but also on which of the two $\varkappa_i$'s is the larger one.
For example, if $\varkappa_1 \gg \varkappa_2$, then the inner angle of the triangle $\delta_1$
always stays small across the ring, while $\delta_2$ changes from zero to $\pi$ as one moves from the outer to the inner
boundary. In this case one needs large $m_2$ and not large $m_1$ in order to produce many interference fringes.
One also observes that the cross section grows towards the boundaries. 

\subsection{Realistic vortex beams}

\begin{figure}[h]
\centering
\includegraphics[width=0.32\textwidth]{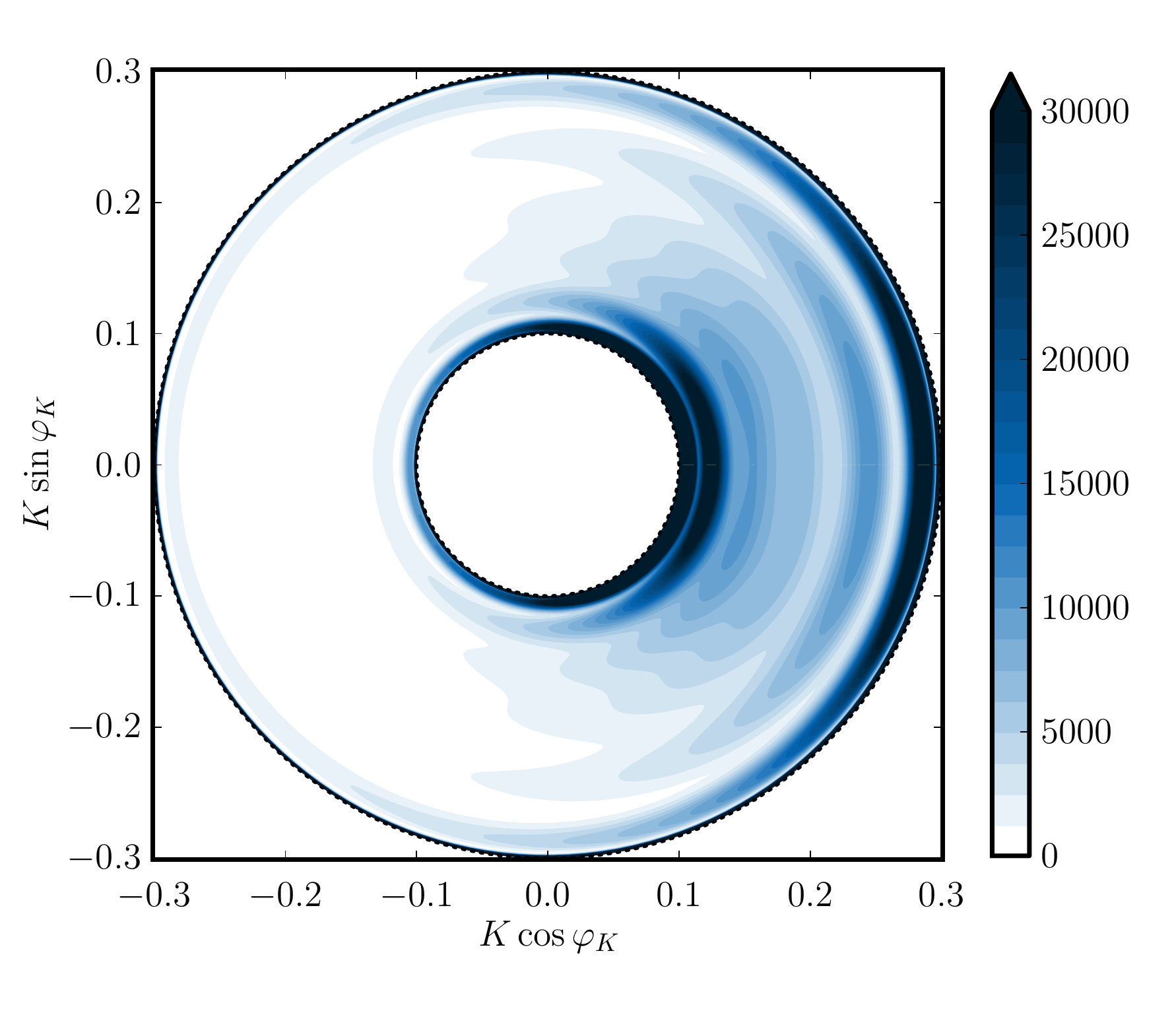}
\hfill
\includegraphics[width=0.32\textwidth]{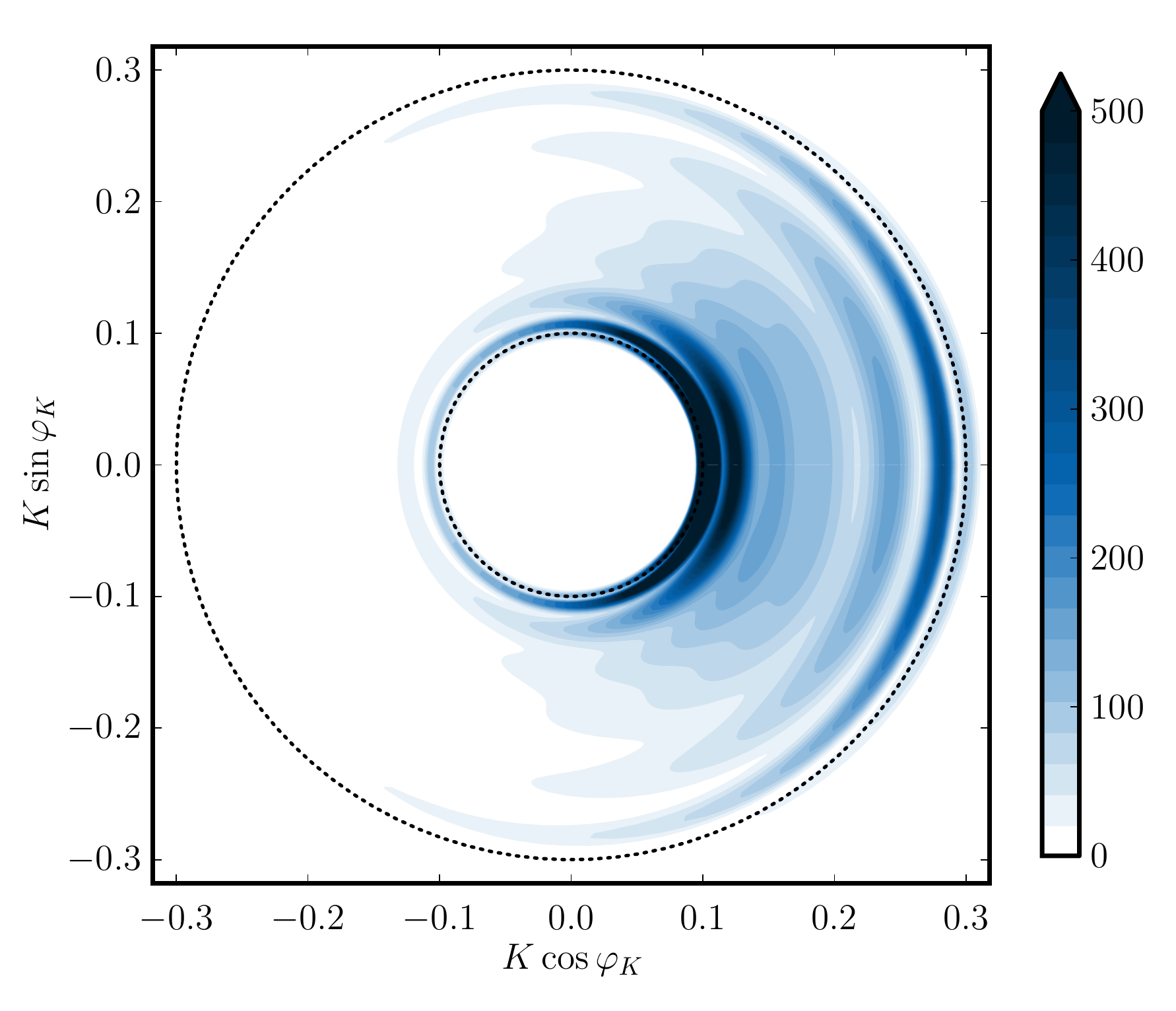}
\hfill
\includegraphics[width=0.32\textwidth]{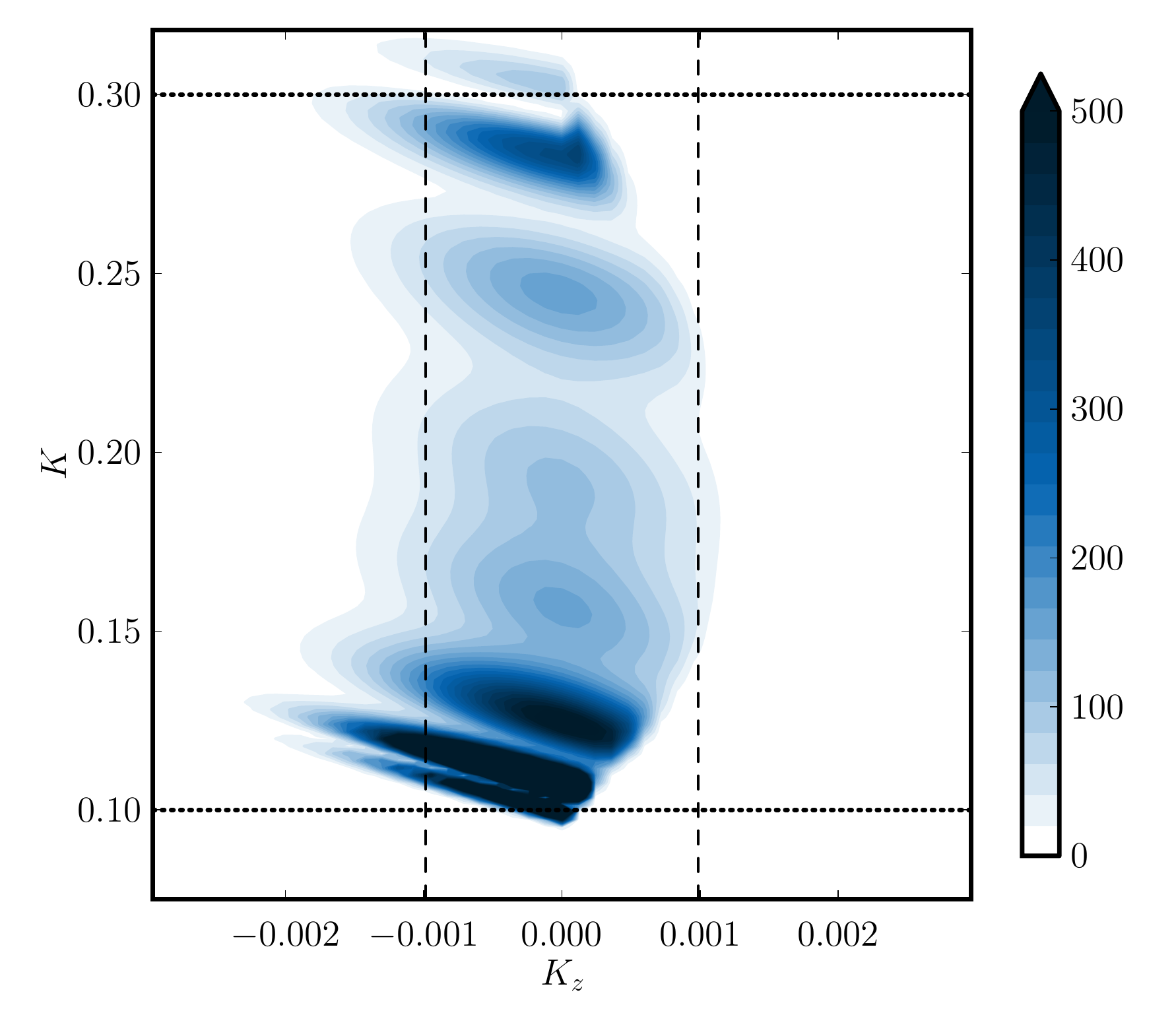}
{\caption{\label{fig-K-plots} 
{\em Left:} Differential cross section for pure Bessel vortex beams, in arbitrary units,
as a function of $\bK'$ for fixed $\bk_1'$ and for the choice of parameters given in (\ref{parameters}).
{\em Middle:}
the same for realistic Gaussian-averaged vortex beams averaged around $\bar\varkappa_i$ with $\sigma_i=\bar\varkappa_i/20$ 
at $K_z=0$. {\em Right:} The $K_z$-distribution of the interference fringes.}}
\end{figure}

As we mentioned above, the vortex amplitude squared $|{\cal J}|^2$
diverges at the boundaries of the transverse momentum integration region due to the $1/\Delta^2$ factor.
This divergence is not integrable. For plane-wave scattering,
a similar divergence appears if the transverse momentum delta-function is squared:
$|S_{\rm PW}|^2 \propto [\delta^{(2)}(\bk_1+\bk_2-\bk_1'-\bk_2')]^2$.
The standard remedy against such divergences in the event rate is to regularize the expressions
by calculating them in large but finite volume
and then to divide the event rate by the corresponding flux, which also displays a similarly divergent behavior. 
The obtained ratio is called the (generalized) cross section,
and it stays finite in the infinite-volume limit. 
In the case of pure Bessel beams, the integration over the available phase space behaves
as $\log R$, the radius of the quantization volume. 
The flux is proportional to ${\cal I}$ given by (\ref{I-flux})
which behaves in the same way. Thus, the cross section
remains finite in the infinite-$R$ limit even for pure Bessel states \cite{JS-2011,ivanov-2011,ivanov-2012}. 

In practice, however, the formal remedy above is not needed in realistic situations. 
The finite transverse extent of the incoming vortex electrons, which is much smaller than $R$, provides a natural regularization.
To incorporate it, we model initial electrons with a gaussian-averaged Bessel state: 
\be
\Psi(x) = \int d\varkappa f(\varkappa)\Psi_{\varkappa m k_z \lambda}(x)\,,\quad 
f(\varkappa) \propto \sqrt{\varkappa} \exp\left[-{(\varkappa-\bar\varkappa)^2 \over 2\sigma^2}\right]\,.\label{smeared}
\ee
Notice that the electron in this state is still monochromatic and has the same $m$ and $\lambda$ 
as each pure Bessel electron in the superposition.
The monochromaticity is achieved by varying $k_z$ in accordance with $\varkappa$.
Alternative profiles $f(\varkappa)$ can also be used;
the exact choice eventually depends on the experimental realization of vortex electrons, 
but it has little effect on our conclusions.

Averaging a monochromatic Bessel vortex state over $\varkappa$ induces an averaging over 
a region of longitudinal momentum.
To take it into account, we return to the general expression (\ref{Stw}) and define the vortex amplitude as
\be
\lr{{\cal J}} = \int d\varkappa_1 d\varkappa_2 f_1(\varkappa_1) f_2(\varkappa_2) \delta(k_{1z} + k_{2z} - K_z) 
{{\cal J} \over \sqrt{\varkappa_1\varkappa_2}}\,,\label{J-smeared}
\ee
where the pure Bessel vortex amplitude ${\cal J}$ is given by (\ref{J2}). 
Notice that longitudinal momenta of the two incoming particles
\be
k_{1z} = \sqrt{E_1^2 - m_e^2 - \varkappa_1^2}\,,\quad k_{2z} = - \sqrt{E_2^2 - m_e^2 - \varkappa_2^2}\,,
\ee
are not necessarily equal. In the paraxial approximation, their sum is very small,
so that $\lr{{\cal J}}$ is strongly peaked at $K_z=0$. 
A similar modification takes place for the flux, which we do not describe in detail
because we focus here on the angular distribution and not on the absolute value of the cross section.

The angular distribution for collision of realistic vortex beams is now given by
\be
d \sigma \propto \int |\lr{{\cal J}}|^2 dK_z \cdot \, d^2\bk_1' d^2\bk_2' \,,
\ee
where we explicitly indicate integration over the narrow $K_z$ region centered at zero
and having width $\sim \sigma \varkappa/k_z$.
\begin{figure}[h]
\centering
\includegraphics[width=0.6\textwidth]{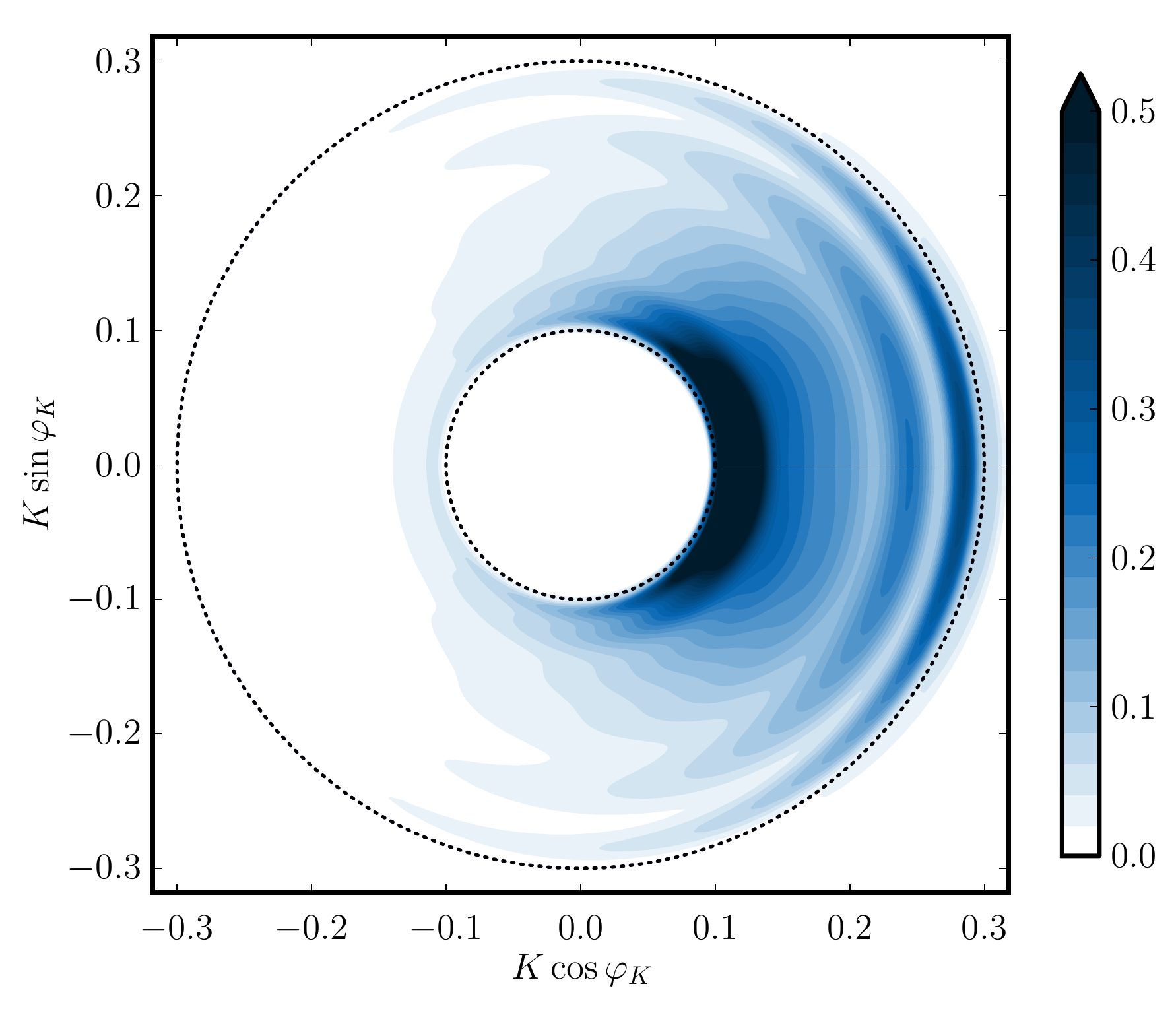}
{\caption{\label{fig-smeared} The $\bK'$-distribution of the M\o{}ller scattering cross section, integrated over $K_z$, 
for realistic vortex beams and for the same set of parameters as in Fig.~\ref{fig-K-plots}.}}
\end{figure}

In Fig.~\ref{fig-K-plots} and Fig.~\ref{fig-smeared}, we present the numerical results for this scattering set-up.
The incoming vortex electrons are averaged here with the Gaussian profile with $\sigma_i = \bar\varkappa_i/20$,
while other parameters are the same as in (\ref{parameters}).
In the middle panel of Fig.~\ref{fig-K-plots} we show the averaged distribution at fixed $K_z = 0$, which exhibits
very similar interference fringes as for the pure Bessel case shown in the left panel of Fig.~\ref{fig-K-plots}
but with the boundary divergences removed.
Notice that the middle panel of Fig.~\ref{fig-K-plots} is not a straightforward averaging of the left panel, 
because weighting with the gaussian profile function is done at the level of amplitude, not the cross section.
This picture depends on $K_z$ as shown in the right panel of
Fig.~\ref{fig-K-plots} in the $(K_z, K)$-plane with $\phi_{K'} = \phi_{1}'$.
As $\varkappa_i$ vary, the $\bK'$-ring
shrinks and expands in the way which is correlated with the total longitudinal momentum $K_z$.
Positive $K_z$ corresponds to $k_{1z} > |k_{2z}|$,
which occurs when $\varkappa_1$ deviates down from $\bar\varkappa_1$ by $\sigma_1$
and $\varkappa_2$ deviates up from $\bar\varkappa_2$ by $\sigma_2$. The inner boundary of the 
ring, $\varkappa_1 - \varkappa_2$ with our choice of parameters, decreases by 
$\sigma_1 + \sigma_2$. As the result, the interference fringes are expected to be oblique 
with the angle $\Delta K_z/\Delta K$ of the order of the opening angle of incoming vortex electrons, 
$\varkappa/k_z$.

We do not require the electron detectors in real experiment to fully reconstruct the final momenta in 3D, 
we just assume that the angular distribution will be measured.
To give predictions for this case,
we need to integrate over $K_z$. It results in a somewhat reduced but still sufficiently high contrast 
of the interference fringes, see Fig.~\ref{fig-smeared}. It is this distribution that 
can be observed experimentally and which we shall further discuss in Section~\ref{section-discussion}. 

\section{Accessing the Coulomb phase}\label{section-coulomb-phase}

In addition to the interference in the angle-differential cross sections, M\o{}ller scattering of vortex electrons 
gives also access to a quantity which cannot be measured in the usual plane-wave scattering.
This is the phase of the (complex) plane-wave scattering amplitude or, more precisely, 
how the phase of the complex scattering amplitude depends on the scattering angle. 

\subsection{The Coulomb phase and its role in particle scattering}

In the one-photon exchange approximation, 
the Coulomb scattering amplitude is purely real, up to inessential phase factors that are
related to the definition of the incoming and outgoing wave functions as in (\ref{Born-UR}).
Higher-order virtual corrections due to multi-photon exchanges give rise
to an imaginary part of the amplitude and after exponentiation
produce the phase $\zeta$. Then, the plane-wave scattering amplitude becomes complex, 
${\cal M} = |{\cal M}| e^{i\zeta}$, and both $|{\cal M}|$ and $\zeta$ depend on the scattering kinematics
(the energy and the scattering angle $\theta$, or the invariant variables $s$ and $t$).

Within the quantum-mechanical treatment of pure Coulomb scattering, 
this extra phase shift arises from the long-range nature of the electromagnetic interactions, 
which distort the incoming and outgoing waves even 
at large distances. One can obtain the 
exact solution for the outgoing wave with a phase shift which grows logarithmically with the separation and 
which depends on the scattering angle \cite{LL}. This behavior is found also in quantum-electrodynamical calculation
of the imaginary part of the two-photon exchange diagrams \cite{Rolnick-1966,west-yennie-1968}.
Due to the infrared (IR) divergence, one usually has to regularize the calculation with a finite photon mass $m_\gamma$, 
the Coulomb phase diverging at $m_\gamma\to 0$. At fixed $m_\gamma$, it displays logarithmic dependence 
on the small scattering angle $\theta$: 
\be
\zeta = \zeta_0(m_\gamma) + 2 \alpha_{em} \eta \ln (1/\theta)\,,\label{zeta}
\ee
where, for electron-electron scattering, 
\be
\eta = \sqrt{1+ {4m_e^4 \over s(s-4m_e^2)}} = 1 + {(1-v)^2 \over 2v}\,,\label{eta}
\ee
and where $v$ is the electron velocity in the center of motion frame.
For ultrarelativistic scattering $v \to 1$ and $\eta \approx 1$, but it grows when $v \ll 1$.
For example, $E_{kin} = 10$ keV corresponds to $\eta \approx 2.6$.
The above expression refers to the Coulomb phase for the scattering of two particles with the same (elementary) electric charge; 
one has to change the sign for opposite charges.

For elastic scattering of charged hadrons, the role of this Coulomb phase becomes more important.
In this case, the elastic scattering amplitude receives contributions from the strong
and electromagnetic interactions, ${\cal M} = {\cal M}_s + {\cal M}_{em}$.
The strong amplitude, together with its own phase which can be very large, is usually poorly known and is the subject of investigation.
Therefore, one wishes to know the Coulomb phase of ${\cal M}_{em}$ as accurately as possible in order to 
probe the unknown strong phase via the interference between the two contributions.

The task of extracting the strong phase via this effect is further complicated by the fact 
that the influence of strong and electromagnetic interactions cannot be fully separated. 
The strong amplitude receives multi-photon corrections, which show IR divergence, 
and it is only the phase difference $\zeta_{em}-\zeta_s$ which is IR finite.
The electromagnetic amplitude calculated at high orders of perturbation series
involves intermediate excited hadronic states, and therefore 
its phase depends on how these states are modelled.

Calculation of the Coulomb phase sparked debates in 1960's, which sometimes reverberate even today.
The first calculation of this phase in small-angle elastic proton-nucleus collision was undertaken by Bethe in \cite{Bethe-1958},
in the potential approach and within the WKB approximation. 
Similar results were obtained later by other authors \cite{rix-thaler-1966,islam-1967}.
Another calculation \cite{solovyev-1966} confronted these results, and the controversy
was resolved by West and Yennie \cite{west-yennie-1968} with the direct diagrammatic calculation.
In later works, more refined calculations of the Coulomb phase were performed \cite{cahn-1982,selyugin-1999,kopeliovich-2001},
and these expressions were used to gain novel insights into the electron-nucleus
deep-inelastic scattering and the elastic small-angle $pp/p\bar p$ scattering \cite{block-1996,kopeliovich-2001b,prokudin-2002}.
More discussion on the role of the Coulomb phase on extraction of the strong interaction amplitudes
can be found in the recent review \cite{dremin-2013}.
This long history shows that the Coulomb phase is an important quantity 
which has received significant attention and which is needed for a safe interpretation of various hadronic processes.

\subsection{Extracting the Coulomb phase}

Despite its importance, the Coulomb phase has never been measured directly in any scattering experiment.
This cannot surprise since only the cross section $d\sigma \propto |{\cal M}|^2$ is available 
in elastic scattering of two plane waves and this renders the phase unobservable.
Here, we use the interference between the two plane-wave amplitudes ${\cal M}_a$ and ${\cal M}_b$
contributing to the elastic scattering of Bessel electron states in order to probe this elusive quantity.
The interference term in the cross section 
\be
d\sigma_{int} \propto 2|{\cal M}_a||{\cal M}_b|\cos(2m_1\delta_1 + 2m_2\delta_2 + \zeta_a - \zeta_b)
\ee 
produces interference fringes, whose pattern is sensitive to the {\em phase difference}:
\be
\zeta_a - \zeta_b = 2\alpha_{em}\eta \ln{\theta_b \over \theta_a} 
\approx \alpha_{em}\eta \ln{t_b \over t_a} \,.
\ee 
This phase difference can be extracted with the aid of the transverse asymmetry $A_\perp$ as defined in (\ref{asymmetries}).
Let us show this procedure in the simplified case of ultrarelativistic small-angle scattering 
with pure Bessel beams described in Section~\ref{section-qualitative}.
Neglecting the small higher-order QED corrections to the modulus of the Born amplitude (\ref{Born-UR}),
we just multiply it with the phase factor $e^{i\zeta(t)}$.
Keeping track only of the cross section dependence
on the final state momenta, we express the cross section as
\be
{d\sigma \over d^2\bk_1' d^2\bK'} = {C \over \Delta^2} \left[{1 \over t_a^2} + {1 \over t_b^2} + 
{2 \over t_a t_b}\cos(2m_1\delta_1+2m_2\delta_2 + \zeta_a-\zeta_b)\cos\delta_1\cos\delta_2\right]\,,\label{Jsquared3}
\ee
with the same factor $C$ as in  (\ref{J2tw2}).
The Coulomb phase is small due to small $\alpha_{em}$,
which allows us to express $d\sigma$ as the Born-type cross section (\ref{J2tw2})
together with a small correction:
\bea
&&{d\sigma \over d^2\bk_1' d^2\bK'} = {d\sigma_{\rm Born} \over d^2\bk_1' d^2\bK'} - {C \over \Delta^2} 
{2 \over t_a t_b}\sin(2m_1\delta_1+2m_2\delta_2)\cos\delta_1\cos\delta_2\cdot (\zeta_a-\zeta_b)\nonumber\\
&&\qquad= {d\sigma_{\rm Born} \over d^2\bk_1' d^2\bK'} - {C \over \Delta^2} {8\alpha_{em}\varkappa_1 \over |\bk_1'|^5}
\sin(2m_1\delta_1+2m_2\delta_2)\cos\delta_1\cos\delta_2 \sin\delta_1 \sin(\phi'_1-\phi_{K'})\,. \label{Jsquared4}
\eea
In the last step of the derivation, we used Eq.~(\ref{tatb}) and the simplifying assumption $\varkappa_i \ll |\bk_i'|$.
The extra term contains $\sin(\phi'_1-\phi_{K'})$, which gives rise to a non-zero transverse asymmetry $A_\perp$
(\ref{asymmetries}).
For realistic electron vortex beams, $A_\perp \sim \alpha_{em}\varkappa_1/|\bk_1'|$.
The exact value strongly depends on the gaussian averaging procedure as well as on the parameters
of the vortex electrons.
Note that for purely real scattering amplitudes, the cross section cannot contain the $\sin(\phi'_1-\phi_{K'})$.
Therefore, the measurement of a non-zero $A_\perp$ will reveal the desired phase difference and
will allow one to reconstruct the Coulomb phase as a function of $t$.

\subsection{Numerical results}

\begin{figure}[h]
\centering
\includegraphics[width=0.40\textwidth]{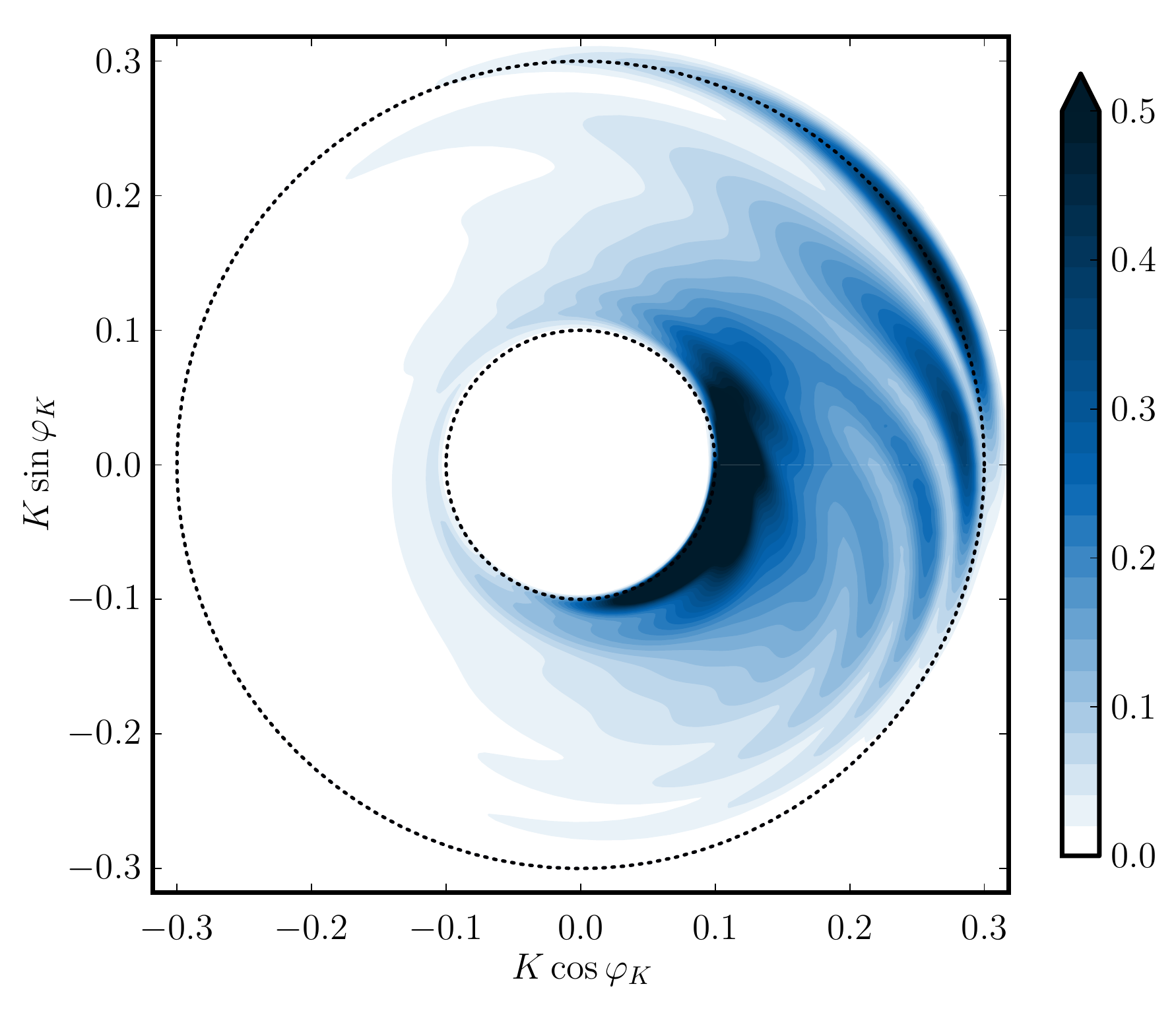}
\hfill
\includegraphics[width=0.58\textwidth]{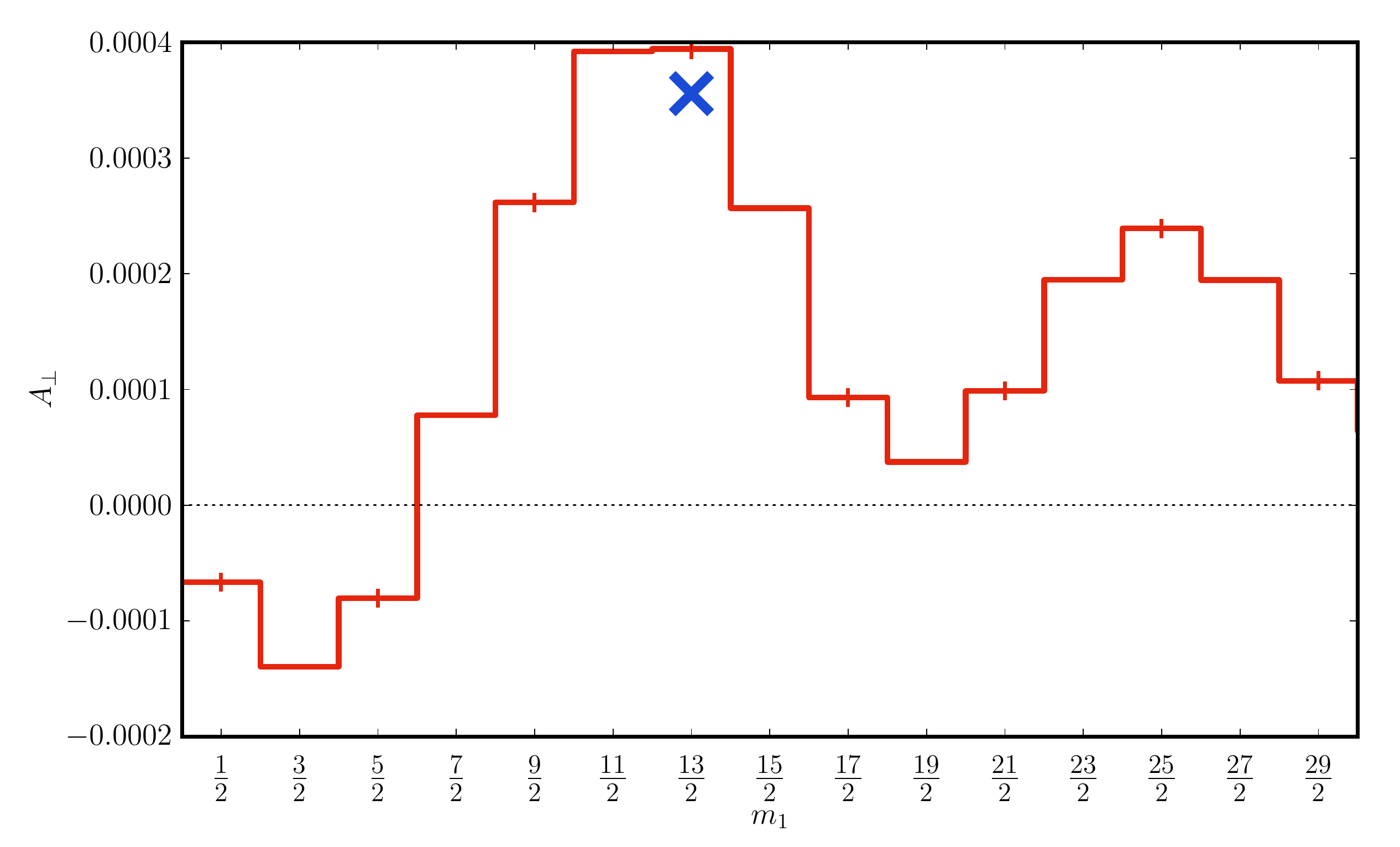}
{\caption{\label{fig-distorted} {\em Left:} Differential cross section, in arbitrary units,
as a function of $\bK'$ for fixed $\bk_1'$.
Here, the same parameters are applied as in Fig.~\ref{fig-smeared}
and, for the sake of illustration, the Coulomb phase prefactor
is artificially set $\alpha_{em}\eta = 10$; 
{\em Right:} asymmetry $A_\perp$ for $m_2 = 3/2$ and various values of $m_1$ 
for the physical value of $\alpha_{em} = 1/137$ and $K_z=0$. The cross indicates the asymmetry for the 
full $K_z$ integration.}}
\end{figure}

Now we return to the exact expressions in general kinematics and
again take the plane-wave scattering amplitude as the Born-level amplitude multiplied 
by the Coulomb phase factor.
The left panel of Fig.~\ref{fig-distorted} illustrates how the $\bK'$-distributions are modified by inclusion
of the $t$-dependent Coulomb phase.
Here we used the same parameters as for Fig.~\ref{fig-smeared}.
In this plot we artificially set the prefactor $\alpha_{em}\eta$ to $10$ 
instead of its typical value ${\cal O}(10^{-2})$ to make its effect more visible.
The obtained large phase clearly manifests 
itself via the strongly distorted pattern of the interference fringes.
For the physical values of $\alpha_{em}\eta$, the up-down asymmetry is not directly visible
and one needs to accurately measure the asymmetry $A_\perp$ in order to detect it.
By adjusting initial parameters, we can optimize this asymmetry further.
The right panel of Fig.~\ref{fig-distorted} displays the results for $A_\perp$ with $\alpha_{em} = 1/137$ and $\eta=1$
as a scan over values of $m_1$ keeping $m_2$ fixed at $3/2$, other parameters remaining as above.
To optimize computer running time, these values are calculated for $K_z=0$. 
Working with the fully $K_z$-integrated distributions does not change the result significantly;
the cross shows one example of such calculation.

One sees that the asymmetry stays $A_\perp = {\cal O}(10^{-4}-10^{-3})$.
This effect is small, mostly due to the smallness of $\alpha_{em}$, but can be detected.
It may be enhanced further with the alternative definition 
of the unpolarized electron.\footnote{With our definition, the asymmetry $A_\perp$ integrated over the entire $\bK'$-ring suffers 
from the partial cancellation between the local asymmetries in the inner and outer parts of the ring.
This is due to the $\cos\delta_1\cos\delta_2$ factor appearing in front of $\sin(\phi'_1-\phi_{K'})$ in (\ref{Jsquared4}).
If the unpolarized Bessel electrons are produced in a state of definite OAM $\ell$ 
rather than the total AM $m$, this factor would be absent.}
Also, by adjusting parameters, one can find a kinematical configuration which would be more sensitive to the phase difference.
Here, we do not attempt a detailed analysis of this phase difference as the exact relation between
$A_\perp$ and phase difference will depend on the details of experiment. 

\section{Feasibility of the proposed experiments}\label{section-discussion}

The proposed experiment can be realized with present day beams and detectors. 
Vortex electron beams with kinetic energies up to 300 keV were created several years ago \cite{twisted-electron}
and helped reveal novel features of how electrons behave in external magnetic fields \cite{thesis,probing,monopole}.
Scattering of two vortex electron beams has not yet been studied experimentally,
but it can be readily done once the instrumentation is modified for this purpose.
Vortex beams can be focused to the focal spot of radius $r \sim $ 1 \AA\, \cite{angstrom},
and high contrast images in these experiments
suggest that stable alignment of the two beams within the common focal spot can be achieved.
The potentially detrimental effects of misalignment are discussed in Appendix~\ref{appendix-misalignment}.
In short, if the shift of the two axes is smaller than the size of the focal spot and if their tilt is smaller
than the vortex beam opening angle, the interference fringes will persist.

Let us now make a rough estimate of whether the present day instrumentation is capable
of realizing the proposed experiments.
M\o{}ller scattering cross section is of the order of $4\pi\alpha_{em}^2/|t|_{\rm min} \sim 4\pi r_e^2/\theta_{min}^2$,
where $r_e \approx 3\cdot 10^{-15}$m is the classical radius of the electron, see Eq.~(\ref{dsdt}).
In the regime where electrons are produced, focused, and collide one by one, 
the probability of such a collision $P$ is given by the cross section divided by the area of the focal spot,
which gives $P \sim [r_e/(r\theta_{min})]^2$. Modern detectors can detect electrons scattered at small angles.
To be on the conservative side, we take $\theta_{\rm min} \sim$ few degrees,
which leads to $P \sim 10^{-6}$.

Next, with the current of 1 nA easily achievable in the electron microscopes producing vortex electrons,
we get about $10^{10}$ electrons per second and, thus, about a thousand detectable collisions per second.
Each collision can be accurately reconstructed with modern detectors,
whose efficiency is taken to be 30\%. In total, we expect that 
scattering events can be detected at the rate of hundreds Hz. Within a few hours of observation time,
a million-event statistics can be accumulated. 

The electron detectors are supposed to detect pairs of scattered electrons in coincidence
and to accurately measure their angular distribution. 
The detectors do not have to cover large solid angle. They can be of the form of 
annular end-caps covering $2\pi$ of the azimuthal angle and a few degrees 
in the polar angles. The angular resolution must be sufficiently high to reconstruct the ring
structure in the transverse momentum space and interference fringes in it.
For example, for $\varkappa_i \sim $ several keV and $E_{kin} = 300$ keV, 
which are already available, the angular resolution of $10^{-3}$ will be sufficient.

The experiments could be carried out as follows.
One prepares two vortex electron beams and brings them in collision in a common focal spot.
When the colliding electrons scatter, they are detected in coincidence by the electron detectors.
For each collision event, the detectors reconstruct the final momenta $\bk_1'$ and $\bk_2'$.
A sufficiently large statistics of such events
can be sliced into several $|\bk_1'|$ regions and, in each region,
one can reconstruct its distribution over $\bK'$.
According to our calculations, interference fringes should be well visible within the $\bK'$ ring
with the million-event statistics.

In order to detect the non-zero asymmetry $A_\perp$ and to probe the Coulomb phase,
the distributions in the annular $\bK'$-region need to be measured with much higher accuracy,
which seems to be challenging with the present day instrumentation.
Once the fringes are detected in a proof-of-principle experiment, 
one can look for ways to improve the set-up.

\section{Conclusions}

In this paper, we presented detailed quantum-electrodynamical calculation 
of the elastic scattering of two vortex electron beams.
We developed the formalism based on the exact description of relativistic vortex electrons
and accompanied calculations with detailed qualitative discussion and numerical results. 

We showed that this process serves as the momentum-space analogue of the classical Young's double-slit experiment.
It reveals interference between two well localized ``paths'' in momentum space,
that is, two plane-wave scattering amplitudes with different momentum transfers.
This interference leads to intensity fringes, which can be detected with present day technology.
As a non-trivial application of this momentum-space interferometry, we suggested to directly measure 
the momentum-transfer dependence of the Coulomb phase, i.e. the phase factor that accompanies all charged particle scattering.
Despite being under theoretical debates and playing important role in elastic scattering of hadrons,
this quantity has never been measured experimentally. We show that elastic scattering of vortex electrons 
gives access to this quantity. 
None of these effects can be measured in the traditional collisions experiments.

\bigskip

The work of I.P.I. was supported by the Portuguese
\textit{Fun\-da\-\c{c}\~{a}o para a Ci\^{e}ncia e a Tecnologia} (FCT)
through the FCT Investigator contract IF/00989/2014/CP1214/CT0004
under the IF2014 Programme, as well as
under contracts UID/FIS/00777/2013 and CERN/FIS-NUC/0010/2015,
which are partially funded through POCTI, COMPETE, QREN, and the EU.
I.P.I. is also thankful to Helmholtz Institut Jena for hospitality during his stay
as a Visiting Professor funded
by the ExtreMe Matter Institute EMMI, 
GSI Helmholtzzentrum f\"{u}r Schwerionenforschung, Darmstadt. 
S.F. acknowledges support by the QUTIF priority programme of the DFG (FR 1251/17-1).

\appendix

\section{Scattering of wave-packets}\label{app-packet}

\subsection{Exact expressions}
The general theory of scattering of non-monochromatic, arbitrarily shaped,
partially coherent beams was developed in \cite{KSS-1992} in terms of Wigner distribution.
For the specific case of pure, monochromatic, and approximately paraxial initial states this formalism can be simplified \cite{ivanov-2012}.
Here we briefly review it for the sake of completeness.

We consider two-particle scattering and assume that the initial particles are described with the coordinate wave functions
$\psi_1(\vec r)$ and $\psi_2(\vec r)$ normalized by $\int d^3 r |\psi_i(\vec r)|^2 = 1$. 
If the wave function is normalizable, the integral here extends to the entire space.
If not, it goes over a large but finite quantization volume $V$, and one needs to check that 
the cross section is independent of $V$.
The corresponding momentum-space wave functions are
\be
\varphi(\vec k) = \int d^3 r\, \psi(\vec r)\, e^{i \vec k \vec r}\,,\quad
\int {d^3 k \over (2\pi)^3} |\varphi(\vec k)|^2 = 1\,.
\ee
The $S$-matrix element for elastic scattering of this initial state into the plane-wave final state with momenta $k'_1$ and $k'_2$
can be written as
\be
S = \int {d^3 k_1 \over (2\pi	)^3} {d^3 k_2 \over (2\pi)^3} \varphi_1(\vec k_1) \varphi_2(\vec k_2) S_{PW}\,,
\ee
where the plane-wave $S$-matrix element $S_{PW}$ is given by (\ref{SPW}).
Since the beams are monochromatic,
the number of scattering events into a given differential volume of the final phase space per unit time is
\be
d\nu = {(2\pi)^7 \delta(E) \over 4 E_1 E_2} \, |F|^2 \, {d^3 k'_1 \over (2\pi)^3 2E'_1} {d^3 k'_2 \over (2\pi)^3 2E'_2}\,.\label{dnu1}
\ee
Here $\delta(E)$ stands for $\delta(E_1+E_2-E'_1-E'_2)$, and
\be
F =  \int {d^3 k_1 \over (2\pi)^3} {d^3 k_2 \over (2\pi)^3}  \varphi_1(\vec k_1) \varphi_2(\vec k_2) \delta^{(3)}(\vec k_1 + \vec k_2 - \vec K')\, 
{\cal M}(k_1,k_2;k_1',k_2')\,,\label{F}
\ee
with $\vec K' = \vec k_1' + \vec k_2'$.
Note that each $\varphi_i(\vec k_i)$ contains a delta-function of the form $\delta(\vec k_i^2 + m^2 - E_i^2)$
because the initial states are monochromatic. 
Thus, the expression for $F$ includes five delta-functions and six integrations and can be represented
as a one-dimensional residual integral. 

\subsection{Plane wave limit}
Let us now see how (\ref{dnu1}) simplifies in the plane-wave limit. 
This limit corresponds to very compact momentum wave functions $\varphi_i(\vec k_i)$ localized
near $\lr{\vec k_i}$. The matrix element can then be approximated as
${\cal M}(k_1, k_2;k_1',k_2') \approx {\cal M}(\lr{k_1}, \lr{k_2};k_1',k_2') \equiv {\cal M}_0$, and the expression for $F$ becomes
\be
F  = {\cal M}_0 \int {d^3 r \over (2\pi)^3} e^{i\vec K' \vec r} \psi_1(\vec r)\psi_2(\vec r)\,.
\ee
Changing $d^3 k'_1 d^3 k'_2$ to $d^3 k'_1 d^3 K'$ and integrating $|F|^2$ over $\vec K'$, one gets
\be
\int d^3 K' |F|^2 =  |{\cal M}_0|^2 \int {d^3 r\over (2\pi)^3} |\psi_1(\vec r)|^2  |\psi_2(\vec r)|^2\,.
\ee
This means that in the plane wave limit, $k_i = \lr{k_i}$, one can effectively replace
\be
|F|^2 \to  |{\cal M}_0|^2  \delta^{(3)}(\lr{\vec k_1} + \lr{\vec k_2} - \vec K') \int {d^3 r\over (2\pi)^3} |\psi_1(\vec r)|^2  |\psi_2(\vec r)|^2\,.
\ee
This expression exhibits an important feature: 
the amplitude which describes the microscopic dynamics and the parameters of the wave-packet factorize.
The number of events can therefore be split into the cross section and luminosity factors:
\bea
&& d\nu = d\sigma \cdot L\,,\label{dnuPW}\\[2mm]
&& d\sigma = {(2\pi)^4 \delta^{4}(\lr{k_1}+\lr{k_2}-k'_1-k'_2) \over 4 E_1 E_2 v } \, |{\cal M}_0|^2 
{d^3 k'_1 \over (2\pi)^3 2E'_1} {d^3 k'_2 \over (2\pi)^3 2E'_2}\,,\nonumber\\ 
&& L = v \int d^3 r\, n_1(\vec r) n_2(\vec r)\,,\quad n_i(\vec r) \equiv |\psi_i(\vec r)|^2\,.\label{lumiPW}
\eea
Note that we inserted here by hand the relative velocity of the two plane waves, $v = |\vec v_1 - \vec v_2|$.

We stress that the separation of the number of events into the differential cross section and 
the (conventional) luminosity is uniquely defined only for plane waves.
Extending this splitting for non-plane-wave collisions is a matter of convention. One needs
to introduce the notion of {\em generalized cross section} \cite{KSS-1992}, for example,
by dividing the full $d\nu$ in (\ref{dnu1}) by $L$ (\ref{lumiPW}) with $v$ defined for $\lr{\vec k_i}$
rather than $\vec k_i$.
With this definition, the generalized cross section for non-plane-wave scattering
takes form
\be
d\sigma = d\sigma_0\, R\, d^3 K' \,,\quad R \equiv {(2\pi)^3\, |F|^2 \over |{\cal M}_0|^2  \int d^3 r\, n_1(\vec r) n_2(\vec r)}\,,
\label{ratioR1}
\ee
where $d\sigma_0$ is the plane-wave $\vec K'$-integrated cross section.
In this notation, the plane-wave limit corresponds to $R \to  \delta^{(3)}(\vec k_1 + \vec k_2 - \vec K')$.

\subsection{Bessel state limit}

Let us also recover the pure Bessel limit from the general expression and compare it with Section~\ref{section-pureBessel}.
With the kinematics conventions adopted there, this limit corresponds to
\be
\phi_1(\vec k_1) = 2\pi N_{\rm tw} a_{\varkappa_1,\,m_1}(\bk_1) \delta(k_{1z}-k_z)\,,\quad
\phi_2(\vec k_2) = 2\pi N_{\rm tw} a_{\varkappa_2,\,m_2}(\bk_2) \delta(k_{2z}+k_z)\,,
\ee
Then the expression for $F$ is simplified to
\be
F = N_{\rm tw}^2 \delta(K_z) \cdot {(-i)^{m_1+m_2} \over (2\pi)^3\sqrt{\varkappa_1\varkappa_2}} \cdot {\cal J}\,,
\ee
where ${\cal J}$ is given by (\ref{J}). Substituting it into the general formula
(\ref{dnu1}), we recover expression (\ref{dnu0}) from the main text.

\section{Helicity amplitudes for M{\o}ller scattering}\label{app-helicity}

\subsection{Exact expressions}
In the Born approximation, the $ee\to e'e'$ scattering amplitude is \cite{LL}
\be
{\cal M} = {\cal M}_t + {\cal M}_u = e^2\left({\bar u'_1 \gamma^\mu u_1\, \bar u'_2 \gamma_\mu u_2 \over t}
- {\bar u'_2 \gamma^\mu u_1\, \bar u'_1 \gamma_\mu u_2 \over u}\right)\,.
\ee
The helicity amplitudes $\lambda_1 \lambda_2 \to \lambda_1'\lambda_2'$
can be represented in the following way:
\be
{\cal M}_t = {e^2 \over t}\Bigl[(Q_{11}Q_{22} + P_{11}P_{22})[12][1'2']^* + (Q_{11}Q_{22} - P_{11}P_{22})(12^{\prime *})(1^{\prime *}2)\Bigr]\,,
\ee
where
\bea
Q_{ij} &=& \sqrt{(E'_i+m_e)(E_j+m_e)} + (2\lambda'_i)(2\lambda_j)\sqrt{(E'_i-m_e)(E_j-m_e)}\,,\\
P_{ij} &=& (2\lambda_j)\sqrt{(E'_i+m_e)(E_j-m_e)} + (2\lambda'_i)\sqrt{(E'_i-m_e)(E_j+m_e)}\,,\\[1mm]
&& [ij] \equiv w^{(\lambda_i)}_{ia}\epsilon^{ab}w^{(\lambda_j)}_{jb}\,,\quad
(ij) \equiv  w^{(\lambda_i)}_{ia}\delta^{ab}w^{(\lambda_j)}_{jb}\,.
\eea
The contraction of spinors depends on helicities. For example, for positive helicities,
\bea
[i^{(+)}j^{(+)}] &=& c_i s_j e^{-i(\phi_i - \phi_j)/2} - s_i c_j e^{i(\phi_i - \phi_j)/2}\,,\nonumber\\
(i^{(+)}j^{(+)}) &=& c_i c_j e^{-i(\phi_i + \phi_j)/2} + s_i s_j e^{i(\phi_i + \phi_j)/2}\,.\label{basic-products}
\eea
where $c_i \equiv \cos(\theta_i/2)$, $s_i \equiv \sin(\theta_i/2)$.
For other helicity choices, these products can be expressed in terms of those given by Eq.~(\ref{basic-products}).
Everywhere, asterisk means complex conjugation.
Note that half-integer values in front of azimutal angles 
will be compensated by half-integer $m$'s in (\ref{J2}).
In all cases, the angles are defined with respect to the same coordinate frame.  
Finally, the $u$-channel amplitude is
\be
{\cal M}_u = -{e^2 \over u}\Bigl[(Q_{12}Q_{21} + P_{12}P_{21})[21][1'2']^* + (Q_{12}Q_{21} - P_{12}P_{21})(11^{\prime *})(2^{\prime *}2)\Bigr]\,,
\ee
with the same definitions as before. Notice that since $[21] = - [12]$, the $[12][1'2']^*$ terms add up 
in ${\cal M}_t$ and ${\cal M}_u$.

\subsection{Ultrarelativistic limit}

In the ultrarelativistic limit, $Q_{ij} = 2 \sqrt{E'_i E_j} \delta_{\lambda'_i,\lambda_j} = (2\lambda_j) P_{ij}$,
so that helicities are always conserved along each fermion line.
The $t$ and $u$ channel amplitudes become
\bea
{\cal M}_t &=& {e^2 \over t}\, 8\sqrt{E_1E_1'E_2E_2'} \delta_{\lambda'_1,\lambda_1}\delta_{\lambda'_2,\lambda_2}
\Bigl(\delta_{\lambda_1,\lambda_2}[12][1'2']^* + \delta_{\lambda_1,-\lambda_2}(12^{\prime *})(1^{\prime *}2)\Bigr)\,,\\
{\cal M}_u &=& -{e^2 \over u}\, 8\sqrt{E_1E_1'E_2E_2'} \delta_{\lambda'_1,\lambda_2}\delta_{\lambda'_2,\lambda_1}
\Bigl(-\delta_{\lambda_1,\lambda_2}[12][1'2']^* + \delta_{\lambda_1,-\lambda_2}(11^{\prime *})(2^{\prime *}2)\Bigr)\,,
\eea
Squaring them, summing over final and averaging over initial helicities, and observing that 
\bea
s &=& 4E_1E_2 \bigl|[12]\bigr|^2 = 4E'_1E'_2 \bigl|[1'2']\bigr|^2\,,\nonumber\\
t &=& -4E_1E'_1 \bigl|[11']\bigr|^2 = -4E_2E'_2 \bigl|[22']\bigr|^2\,,\nonumber\\
u &=& -4E_1E'_2 \bigl|[12']\bigr|^2 = -4E_2E'_2 \bigl|[21']\bigr|^2\,.
\eea
we obtain
\be
{1 \over 4}\sum|{\cal M}|^2 = 2 e^4 \left[\left({s \over t} + {s \over u}\right)^2 + {u^2 \over t^2} + {t^2 \over u^2}\right] =
2 e^4 {s^4+t^4+u^4 \over t^2u^2}\,,
\ee
which leads to the well-known result \cite{LL}
\be
{d\sigma \over dt} = {2\pi\alpha_{em}^2 \over s^2} {s^4+t^4+u^4 \over t^2u^2}\,.\label{dsdt}
\ee

\section{Imperfect alignment}\label{appendix-misalignment}

In Section~\ref{section-main} we assumed that two colliding vortex electrons are perfectly aligned, 
that is, they can be both described by Eq.~(\ref{bessel}) and with the same quantization axis.
Real beams can be slightly misaligned, either due to shift or tilt between the two axes.
One may wonder whether a misalignment 
can cause a deviation of the angular distribution which could mimic the effect we measure, 
such as the visibility of the interference fringes or the azimuthal asymmetry.
Here, we will briefly discuss its effects without undertaking a full numerical simulation, 
which in any event would heavily rely on the details of experiment.

A shift between two parallel axes can be easily incorporated in the above formalism. 
A vortex state $|\varkappa, m\rangle_a$ defined with respect to an axis
shifted in the transverse direction by vector $\ba = a(\cos\phi_a,\,\sin\phi_a)$
can be expressed via the vortex states defined for the original axis:
\be
|\varkappa, m\rangle_a = \sum_{m' = -\infty}^{+\infty} e^{i(m'-m)\phi_a} J_{m'-m}(\varkappa a) |\varkappa,m'\rangle\,.
\ee
Using this representation for the first electron and working out some algebra, 
one find that ${\cal J}$ in (\ref{J2}) is replaced with
\be
{\cal J}_{\mathrm{shift}}  \propto {\cal M}_{a}\, e^{i\varkappa a \sin\delta_1 \cos\phi_a} e^{i (m_1 \delta_1 + m_2 \delta_2)} 
+ {\cal M}_{b}\, e^{-i\varkappa a \sin\delta_1 \cos\phi_a}e^{-i (m_1 \delta_1 + m_2 \delta_2)}\,.\label{J3}
\ee
Since such a shift effectively makes the scattering amplitude complex, one might worry that
it will induce the azimuthal asymmetry $A_\perp$ mentioned above.
This is not the case. The extra phase factor is opposite in the two plane-wave amplitudes
and it does not depend on $t$.
If the two plane-wave amplitudes are real, the net effect is a shift of the interference fringes without reducing their contrast.
For example, for the ultrarelativistic small-angle scattering considered in Section~\ref{section-qualitative},
this phase factor amounts to the replacement
\be
\cos(2m_1 \delta_1 + 2 m_2 \delta_2) \to \cos(2m_1 \delta_1 + 2 m_2 \delta_2 + 2 \varkappa a \sin\delta_1 \cos\phi_a)
\ee
inside Eq.~(\ref{Jsquared}).
The cross section remains up-down symmetric and no $A_\perp$ is induced.
For Gaussian-averaged beams, the effect of the extra phase factor can be sizable only
if $\sigma a > 1$, that is, if the shift between the two axes is larger than the focal spot.
We expect that the high control over vortex beams will allow for a good overlap 
of the two colliding beams.

A tilt of the two axes has a more important effect on the interference pattern. 
For pure Bessel beams, 
the conservation of energy and momentum will spoil the two-slit picture 
which works for parallel axes. 
However, for physical states smeared over some region of $\varkappa$,
the interference is restored provided the tilt angle is less than $\sigma/k_z$,
that is, the factor of $\sigma/\varkappa$ of the Bessel state opening angle.
This should be well achievable experimentally.
In short, if the two axes are aligned with sufficient accuracy, 
the two-slit interference pattern survives.

\end{document}